\documentclass[twocolumn]{article} 

\usepackage{arxiv}

\usepackage[all]{nowidow}

\usepackage[utf8]{inputenc} 
\usepackage[T1]{fontenc}  
\usepackage{hyperref}       
\usepackage{url}            
\usepackage{booktabs}       
\usepackage{amsfonts}       
\usepackage{microtype} 

\usepackage{float}  
\usepackage{graphicx}
\usepackage[labelfont=bf]{caption}
\usepackage{tcolorbox} 
         
\usepackage{enumitem}                  
\usepackage{enumerate}

\newcommand{\interview}{data engagement interview~}
\newcommand{\interviewNS}{data engagement interview}

\newcommand{\interviews}{data engagement interviews~}
\newcommand{\interviewsNS}{data engagement interviews}
\newcommand{\Interviews}{Data engagement interviews~}

\newcommand{\etal}{et al.~}

\renewenvironment{quote}%
  {\list{}{\leftmargin=0.1in\rightmargin=0in}\item[]}%
  {\endlist}

\title{An interview method for engaging personal data}

\def \shortauthor {Moore et al.}
\author{
  Jimmy Moore \\
  University of Utah\\
    \texttt{jimmy@cs.utah.edu}
    \And
  Pascal Goffin \\
  Asvito Digital AG \\
     \texttt{ppjgoffin@gmail.com}\\
    \And
  Jason Wiese \\
  University of Utah \\
    \texttt{wiese@cs.utah.edu}
    \And
  Miriah Meyer \\
  University of Utah\\
  \texttt{miriah@cs.utah.edu}
}

\begin{document}
\twocolumn[
  \begin{@twocolumnfalse}
    \maketitle

\begin{abstract}
Whether investigating research questions or designing systems, many researchers and designers need to engage users with their personal data. However, it is difficult to successfully design user-facing tools for interacting with personal data without first understanding what users want to do with their data. Techniques for raw data exploration, sketching, or physicalization can avoid the perils of tool development, but prevent direct analytical access to users' rich personal data. We present a new method that directly tackles this challenge: the data engagement interview. This interview method incorporates an analyst to provide real-time personal data analysis, granting interview participants the opportunity to directly engage with their data, and interviewers to observe and ask questions throughout this engagement. We describe the method's development through a case study with asthmatic participants, share insights and guidance from our experience, and report a broad set of insights from these interviews.
\end{abstract}

\vspace{2mm}


\vspace{8mm}

  \end{@twocolumnfalse}
]

\begin{tcolorbox}[floatplacement=!b,float,left=2mm,right=2mm,top=1mm,bottom=1mm]
\small
This is the authors' preprint version of this paper. License: CC-By Attribution 4.0 International. Please cite the following reference: \\
Jimmy Moore, Pascal Goffin, Jason Wiese, Miriah Meyer.
An interview method for engaging personal data
\textit{Proceedings of the ACM on Interactive, Mobile, Wearable and Ubiquitous Technologies(IMWUT)}, Vol. 5, No. 4, Article 173, December 2021. https://doi.org/10.1145/3494964
\end{tcolorbox}

\section{Introduction} \label{sec:introduction}

Observing how people engage with their personal data offers a wealth of insights for researchers and practitioners. For example, understanding and identifying the kinds of questions people ask of their data, and the analysis strategies they employ to answer them, helps them design new tools \cite{vanKollenburg2018Exploring,schroeder2019examining}. Creating opportunities for people to learn new things from their personal data can also provide triggers for positive behavior changes \cite{Tang2017Harnessing}, and showing participants the value of their personal data can help motivate continued self-tracking \cite{choe2014understanding}. 

As the scope and scale of personal data increases --- through improved sensor resolution and integrating multiple data sources --- 
engaging with data increasingly requires the use of sophisticated analysis tools and methods.
Lightweight approaches, such as sketching \cite{Marasoiu2016} or data physicalizations \cite{Thudt2018}, can be quick to perform and require minimal design effort.  These approaches, however, often involve abstract or incomplete data and do not scale for direct engagement with the complexity of many real-world self-tracked data sets. Examining raw data with these approaches may work for exploring a small amount of data at a time \cite{tolmie2016has}, but can quickly break down with larger data sets. These larger data sets generally require some amount of computation to support exploration and analysis, leading many personal informatics researchers to develop and deploy custom analysis tools in order to engage participants with their data. This heavyweight approach, however, requires significant design work and interpretation of what people might \textit{actually} do from what they \textit{say} they want to do, potentially leading to gaps in analysis support \cite{epstein2014taming}. 

We propose a middle ground approach in this paper that we call the \textit{data engagement interview}. The \interview is a research method that sits between the lighter weight approaches involving minimal design effort, and the more heavyweight approaches that involve customized tool development. We developed data engagement interviews to help researchers better understand and identify what participants want from their personal data by observing participants ask and answer questions in real-time from their own data. This interview method incorporates a dedicated data analyst on the interview team to provide participants with a flexible toolbox of real-time analysis techniques. Using this method, interviewers can support participants as they explore their data to elicit and observe more authentic data engagements, while the data analyst takes direction on how to process or present participants' data to answer personal questions. Whereas \interviews are more resource-intensive than standard interview methods, this method strikes a balance between engagement strategies that fail to incorporate complex personal data and those requiring customized tool development prior to collecting any observations. This interview method can quickly help researchers with eliciting design requirements for potential future system development, while also helping participants use their data to flexibly answer unique and personal questions.
 
We developed the data engagement interview from our own research goals to design new visual analysis tools for asthmatic families living with indoor air quality sensors \cite{Moore2018}.  Through sensor deployments with six households, we collected various data sets for each family that included several months of quantitative and qualitative data, sampled over different timescales and measurement intervals, that require both personal annotations and contextual knowledge to productively interpret and analyze. These computational and contextual demands prevented us from using lightweight engagement methods. After developing the data engagement interview method, we conducted interviews with our participating families to observe how and why they engage with their personal indoor air quality data. In addition to extracting design requirements for a future analysis tool, our analysis of the interview transcripts showed that \interviews can also yield a host of other insights and opportunities.  

The contribution of this work is a framework for conducting \interviewsNS. This framework allows researchers and practitioners to engage participants directly with their personal data without the need to develop custom data analysis tools. We also conduct a case study in which we apply the interview framework to characterize the motivations and analysis tasks of asthmatic families when working with personal air quality information. We observed evidence that this method can expose differences between what participants say they want to do with their data, and what they actually do; engage participants more readily than standard interview methods; teach participants new things about their data;  teach researchers new things about design requirements; and benefit research outcomes by improving insights on study design and motivating participants to self-track.   To support transferability, we have prepared an online guide\footnote{\url{https://vdl.sci.utah.edu/EngagementInterviews}} \cite{dataEngagementInterviewGuide} that includes a sample interview protocol based on our experience of conducting \interviewsNS, along with other detailed suggestions, interview materials, and example data and processing scripts.

Our analysis of participants' \interviews lends evidence that this method can be a promising approach for helping researchers and practitioners learn more about the goals and motivations of their target users. We further speculate that \interviews can be a widely applicable research method, suitable across a broad range of personal informatics domains, and scalable to accommodate different types of personal data and high-resolution, multisource data sets. Although this method is not intended as a replacement for more traditional tools, its success at engaging our participants with their data suggests collaborative analysis via an analyst-in-the-loop is a viable alternative for insight generation compared to using customized tools, and an interesting direction for future work that we briefly discuss in this paper, but detail more thoroughly in a companion paper \cite{moore2021gap}.

Section \ref{sec:related_work} provides background on engagement methods and the space for \interviewsNS.  We describe our process for developing the \interview in Section \ref{sec:methods}, outline a framework for conducting them in Section \ref{sec:interview_framework}, and present a case study of how we applied the framework in Section~\ref{sec:case_study}.  Section~\ref{sec:results} describes outcomes from applying this framework with asthmatic participants engaging with their indoor air quality.  We discuss some consequences of this interview method in Section~\ref{sec:discussion}, limitations of the interview framework in Section~\ref{sec:limitations}, and conclude with ideas for future work in Section~\ref{sec:conclusion}.

\section{Background}\label{sec:related_work}

The growth in technology for capturing data about people's everyday, lived experiences has led to an explosion of personal data and a wealth of new insights. Self-trackers are actively collecting data and learning things about their bodies through fitness trackers and sleep devices~\cite{kay2012lullaby,consolvo2006design,consolvo2008activity,epstein2014taming,epstein2016beyond}; about their environments through air quality monitors and utility usage sensors \cite{Moore2018,kim2009inair,kim2010inair,kim2013inair,fang2016airsense,Gupta2010,Campbell:2010:WMS:1864349.1864378}; about their health through digital diaries and nutrition trackers \cite{tsai2007usability,chung2019identifying,chung2017personal}; and about how they spend their time through calendars and social-media trackers \cite{peesapati2010pensieve,lindqvist2011m}. For personal informatics researchers and practitioners, the explosion in available data sets has created myriad opportunities to learn about how and why people engage with personal data \cite{li2010stage,choe2014understanding,epstein2015lived}, and the kinds of behavioral changes this engagement provokes \cite{kersten2017personal}.
These opportunities, however, require engaging participants with their personal data. In this section, we describe the range of engagement methods researchers and practitioners have at their disposal, and argue for \interviews as a middle ground approach.

\subsection{Lightweight methods}

Design literature provides various methods for informing researchers about what or how to build regarding interactive tools or interfaces. Participatory design \cite{schuler1993participatory} is a common approach that invites users to collaborate in the design process to help inform the final result.  This technique can help identify commonly undertaken tasks, or solicit feedback on the ways they may be improved.  These approaches, however, are tailored for collecting insights that inform design \textit{outcomes} rather than deeply understanding ways to productively engage people with their personal data. Understanding how to engage with personal data requires a deep, situated knowledge of people's lives and routines to accurately interpret \cite{tolmie2016has} and can involve collaboration between a data worker and participant to derive insights or offer advice \cite{fischer2016just,Fischer2017dataWork}.
 
Existing tools that visualize personal data typically support data review through simplified interfaces with minimal interactivity. These tools are mostly designed to show data, not to thoroughly analyze it. Tolmie \etal talk homeowners through their personal data  using a  basic time series plot for displaying sensor measurements~\cite{tolmie2016has}. Other researchers provide similar interfaces to end users for exploring how to support people engaging with their personal air quality data  \cite{kim2009inair,kim2010inair,kim2013inair,fang2016airsense,Moore2018}.   
These interfaces help people gain a sense of what their data \textit{is}, but not what it can \textit{do}.  Without the ability to easily modify or change the data's representation and visualization, these interfaces can support only a limited number of data analysis tasks. 

Alternatively, data sketching provides a lightweight method that has people sketch their impressions of data with minimal design effort.  Data sketching removes the barriers to how data can be organized and formatted to promote brainstorming and collaborative workflows \cite{Browne2011, buxton2007anatomy}, storytelling \cite{Lee2013}, and communicating knowledge about data to others \cite{Marasoiu2016}. The process of sketching also improves thinking \cite{buxton2010sketching},  supplements discussion \cite{Walny2011}, and helps clarify ideas about design \cite{buxton2007anatomy}.  Engaging people with sketching helps them externalize their thoughts and ideas about data organization, visualization goals, and any underlying trends or traits they suspect may live within their data \cite{Walny2011,Marasoiu2016}. In this way, sketching can free people to more quickly communicate organizational goals or ideas, especially in the absence of formal design or analysis vocabulary. Sketching often does not incorporate real data, however, and efforts to encode this information, either by hand or through digital tools, can be slow or complicated \cite{Browne2011}.  Instead, sketching can be a useful design component for \textit{imagining} personal data, but it does not suffice for concrete analysis tasks or questions that require engaging personal data directly.

Data physicalization, another lightweight method, helps people explore and communicate data through geometric or physical properties of an artifact \cite{jansen2015opportunities}. Data physicalization has been successfully applied in workshops \cite{Huron2017} and teaching environments \cite{willett2016constructive} to engage people through prepared data sets. Work by Thudt \etal~\cite{Thudt2018} extends this approach to personal contexts, and uses data physicalizations to bring people closer to their personal data in support of self-reflection. Whereas this approach succeeds at deeply engaging people with their personal data, it requires a significant manual effort, and limits the representational accuracy and scope due to its inherent physical constraints \cite{Thudt2018}. Consequently, the nature and scale of many personal data sources prevent physicalizations as a practical analysis strategy.

\subsection{Heavyweight software}

The messy and complex nature of many personal data sets requires some level of wrangling, formatting, and preprocessing, making it  difficult to integrate into general purpose tools, many of which some people already find hard to use in personal contexts \cite{choe2014understanding}.  As an alternative, researchers, practitioners, and quantified self enthusiasts invest significant effort to design and build bespoke tools for people to engage with their data. These tools typically focus on a narrow set of specific or predefined questions, thereby eliminating the need for users to translate their questions into analysis tasks, or to wrangle their data into an appropriate representation \cite{epstein2014taming, huang2014personal}. This approach, however, does not let users explore a broad set of personally relevant questions, nor does it leverage users' rich, situated, and extensive knowledge of what aspects of the data are personally interesting and insightful, and which are not~\cite{choe2017understanding}. The challenge for designers is that people who have never directly engaged deeply with their data may not be able to predict what they want to do. For example, Epstein \etal surveyed 139 people on common tracking goals, motivations, and influences for informing visual and data analysis criteria to evaluate lifelog data \cite{epstein2014taming}. After developing and deploying a tool to support these goals, subsequent evaluations ``did not find any correlations between valued cuts and the reported goals of participants,'' prompting guidance that users should receive several possible designs, versus ``simply [generating] cuts corresponding to stated goals, as that could deprive trackers of potentially interesting discoveries in their data'' \cite{epstein2014taming}. Even when designing customized solutions, personal informatics tools may still struggle to provide flexible analytic capabilities that completely address or anticipate users' needs.

\subsection{A middle ground approach}

The \interview proposed in this paper takes a middle ground approach  by helping researchers identify user needs through directly engaging these users with their personal data before expending the significant design effort to develop a custom tool.
\Interviews are an adaptation of the pair analytics research method  that captures reasoning processes in visual analytics scenarios \cite{Arias-Hernandez2011}.  Pair analytics borrows from protocol analysis and pair programming techniques by joining a subject matter expert and visualization practitioner to collaboratively tackle a relevant analytical task. This approach avoids the cognitive and social loads reported in standard think-aloud applications~\cite{trickett2000dipsy,dickson2000effects,wilson1994proper} by capturing participants' analytical reasoning through a conversational and collaborative problem-solving process.  This approach, however, requires that participants share equal analytical and computational skills to productively work through their given task, which may not always be the case in personal informatics contexts.

We build on the pair analytics approach and incorporate a dedicated data analyst role within the interview team.  Whereas the interviewer role is responsible for engaging the participant and keeping discussion on topic, the data analyst takes analytic direction from the interview participant. Unlike the standard Wizard of Oz approach \cite{kelley1984iterative} where the interview participant unknowingly interacts with an analyst, the \interview brings the analyst to the forefront to gain the collaborative and conversational benefits of pair analytics.  These interviews provide a personalized analysis experience that allows the researchers and participants to deeply engage in the analysis process, and explore personal data through the incorporation of a dedicated data analyst working with flexible analysis tools and the participants' own data. 

\section{Developing the interview framework} \label{sec:methods}

This section outlines how we developed the \interview framework.  We describe the framework in Section \ref{sec:interview_framework}, and give more detailed descriptions and recommendations for performing \interviews in  Section \ref{sec:case_study}.  Section \ref{sec:results} reports on the outcomes of conducting \interviews with our participants.

\subsection{Motivation}

\begin{figure*}
    \centering
    \includegraphics[scale=0.152]{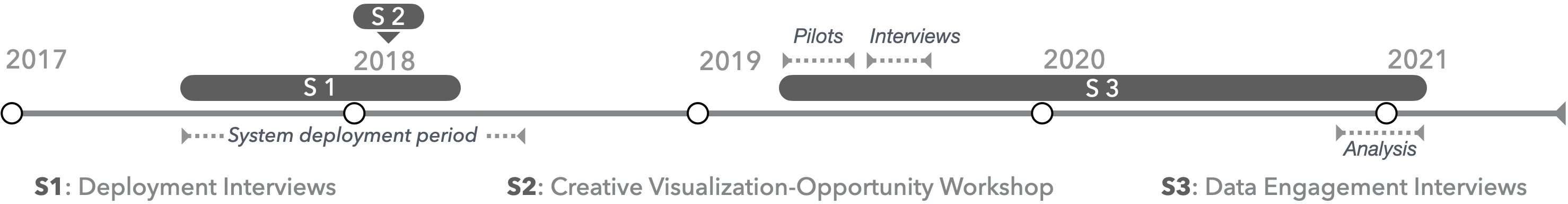}
    \caption{We performed three separate research activities to understand and identify how participants engaged with their personal air quality data. This timeline shows our initial deployment interviews (S1) for tracking how participants engaged with an indoor air quality sensing system \cite{Moore2018}; a participatory workshop (S2) to capture what people wanted to know, see, and do with their indoor air quality data; and \interviews (S3) for observing how participants approached operationalizing their questions and analyzing personal air quality data.}
    \label{fig:research_activity_timeline}
\end{figure*}

We developed the \interview as part of a longitudinal study of people living and interacting with an air quality monitoring system in their homes; \autoref{fig:research_activity_timeline} shows a timeline of the study. In the first study stage (S1), we deployed a system consisting of multiple air quality monitors, mechanisms for residents to annotate their air quality data, and an interactive tablet interface for displaying these measurement data and annotations. We tracked how study participants annotated and interacted with their data through 6 long-term field deployments (20-47 weeks, mean 37.7 weeks) and conducted 3 rounds of traditional in-person interviews with each participant (34 interviews, 20 hours). Our interview data analysis revealed a diverse range of questions the participants had about air quality in their homes, and about the depth of contextual, personal knowledge required to generate insights from their data \cite{Moore2018}.

Following this first stage of research, we had planned to design a visual analysis system to support our participants to more fully engage with their data. The interviews from S1 contained a significant amount of feedback on ways to improve the deployed system's tablet interface; however, further analysis revealed that the suggested improvements would not support the high-level goals participants shared at various points in their deployments. When reflecting on study outcomes in the context of the field deployment,  we understood that our interviews were developed to gauge \textit{how} participants used their air quality system, not \textit{what} tasks they needed to perform in order to answer their personal questions.

To address this shortcoming, we conducted a participatory visualization workshop  \cite{Kerzner2019}  (S2) toward the end of the system deployment period with two participants from S1. The workshop goal was to collect and characterize participants' questions and motivations for hosting an air quality monitoring system in their homes. Combining the data collected in S1 and S2, we again attempted to translate user feedback into design and task requirements for a visual analysis system. We surveyed the range of participants' stated goals, finding that they ranged between the direct and concrete -- \textit{What is the worst time of year for indoor air quality?} -- to  more abstract or out of scope -- \textit{I want product recommendations for improving my air quality}.  The participatory workshop afforded participants an opportunity to critique their previous interface and share retrospective feedback, but it failed to provide insight into types of data analysis tasks that an effective system would need to support. Without direct access to their data, our participatory design approach brought us no closer to understanding what our participants wanted to do, or how they would approach their goals using their data.

To address this question, we needed to provide our participants with a rich, flexible, and accessible set of data analysis techniques, and observe how they would make use of them to answer their personal questions. We developed the \interview as a stand-in for the analysis tool that we did not yet know how to design. 

\subsection{Developing the interview protocol}

In our search for guidance on how we might elicit design requirements from our participants, we found both the visualization and human computer interaction literature lacked any suitable research methods for directly engaging everyday users with their personal data.  We developed \interviews with the assumption that interview participants are \textit{not} analysis experts, and therefore incorporated a dedicated data analyst as an active member in the interview process to offload analysis tasks from the participant. This change helps lower the barrier for engaging with personal data while still providing a rich suite of analysis capabilities. We also incorporated additional ways to elicit participants' analysis goals, such as reviewing physical data printouts and sketching, to help externalize their ideas. Recognizing the potential complexity of the interview dynamics, we further modified our draft protocol by splitting the interviewing responsibilities between two interviewers to maximize our likelihood for collecting and capitalizing on valuable research insights \cite{monforte2021pairinterivew}.  This pair interviewer approach has one interviewer lead the discussion, and the other track the conversational flow to help keep things on task. 

We refined the interview protocol over two rounds of pilot interviews.  The first round of piloting helped streamline and organize the interview structure. We recruited 7 first-round pilot participants from our research lab, 6 of whom were computer science graduate students, and 1 computer science undergraduate student. These first-round pilot interviews did not incorporate a dedicated data analyst. Instead, we had pilot participants role-play as asthmatic self-trackers and sketch what they wanted to do with a set of representative air quality data. 

In the second round of pilot interviews, we incorporated our data analyst into the interview team. We recruited the participants in this pilot study from a convenience sampling of undergraduate students pursuing nonanalytic degrees and nonstudents from an online forum. The second-round pilot participants consisted of 4 dance majors, 1 physical therapist, 1 market researcher, and 1 social media influencer. These pilot interviews focused on evaluating the feasibility of performing real-time data analysis within the interview and improved how we introduced and presented data to participants.   All second-round pilot participants were compensated with a \$20 Amazon gift card.  

\subsection{Study Participants}\label{sec:participants}

We developed the interview framework from our experience conducting seven \interviews with primary participants we retained from S1 \cite{Moore2018}.  These participants were initially selected from a concurrent university-run medical study involving asthmatic families \cite{PRISMS}, and were themselves asthmatic (P1, P2, P4a, P5, P6), or primary caregivers to asthmatic children (P2, P3, P4) --- note that these participant IDs correspond to the more descriptive characterizations provided in previous work \cite{Moore2018}. We included participant P4’s teenage daughter, P4a, as an additional study participant given her significant engagement within this study. No other children actively participated. Spouses of participants P1 and P2, hereafter labeled P1-S and P2-S, also contributed feedback and suggestions during the \interviews but were otherwise not involved in S1 or S2 and not counted among the primary participants. Of the primary participants in our study, 5 were female (P1, P2, P3, P4, P4a) and 2 male (P5, P6). Although we did not collect families' earned income or socioeconomic status, all adult participants had received a high school degree. P2 completed some college education with no degree, and P1, P3, and P4 obtained a college degree. P5 and P6 had master's degrees.  Participants P1 and P2 were stay-at-home mothers, P3 was a web developer, P4 worked as a nurse, P5 was a school administrator, and P6 worked on public policy.   All study participants were compensated with a \$20 Amazon gift card for their participation in the \interviewsNS.

\subsection{Analyzing \interviews}\label{sec:analysis}

The interview framework emerged from analyzing seven \interviews conducted with our study participants. We audio recorded each interview for 8.9 hours of interview audio, and maintained individual Jupyter notebooks in Python from each interview to create a self-documenting record of participants' analysis process and visualizations. These notebooks also allowed us to amass a store of reusable code that we could employ in subsequent interviews \cite{kluyver2016jupyter}. Reviewing and reflecting on these artifacts helped us build an understanding of what our participants wanted to do with their data, and to identify their goals and their overall approach to data analysis.

The interviewers also engaged in reflexive discussion after each \interviewNS, sharing their thoughts and reactions to what each found surprising, unexpected, frustrating, or insightful during an interview. One of the interviewers compiled reflexive notes on these experiences after each interview, which were further supplemented with a secondary summary after listening to the recorded interview audio.  This process captured additional aspects of the interview mechanics, including participants' stated questions, goals, and motivations, while also providing an overall commentary on the interview process. We revisited the reflexive notes frequently throughout the analysis process. 

Each interview audio recording was also professionally transcribed and imported into Google Sheets.  The first author of this paper blocked and summarized individual interview sections of each interview to create a high-level overview summary for other researchers to review.  Three researchers then read through and annotated participant transcripts, and then met to discuss aspects we found noteworthy from a methodological or research perspective for each of the seven interviews. These meetings were also audio recorded. The first author listened to these meeting recordings to further summarize the main discussion points, and populated an affinity mapping board with meeting summary notes.  The resulting affinity diagram was supplemented with direct evidence within the transcripts associated with the summaries. The three researchers involved in analyzing the transcripts then iteratively produced an additional affinity diagram of core interview themes over several days. These themes informed the framework we describe in Section \ref{sec:interview_framework}. The outcomes from applying this interview framework are presented in Section \ref{sec:results} as well as in a companion paper~\cite{moore2021gap}.

\section{Interview Framework}\label{sec:interview_framework}
  
The \interview is a novel interview method that elicits engagement with personal data to support a host of observations and insights about those engagements. This method differs from more traditional interview approaches in two ways: first, through the inclusion of a dedicated data analyst on the interview team who has access to a prepared analytic toolbox and the participant's personal data; and second, by structuring the interview to cycle between exploratory and goal-oriented analysis strategies. The participant directs the analysis in these interviews by communicating their desired analytic tasks to the data analyst, who then performs the requested analysis on the participant's data.  This process allows the participant to engage with their data in a flexible and personally relevant way, and provides researchers and practitioners opportunities to observe what participants \textit{actually} do with their data when given the freedom and resources to do so.
    
\Interviews foster a conversational dynamic around personal data by offloading the analytic burden from the participant, so they can more readily share thoughts on their process, justifications, and reactions, in their own words, as part of a naturally unfolding conversation. Maintaining this conversational dynamic can avoid post hoc rationalizations that can be common to other concurrent or retrospective verbal reports \cite{fonteyn1993description}.  This approach can also help researchers collect more rich and authentic insights into participants' motivations and problem-solving processes Whereas several think-aloud techniques exist in the interviewing literature~\cite{ lenzner2016cognitive, fonteyn1993description, presser2004methods}, few are tailored for specifically engaging visual analytic tasks and processes \cite{Arias-Hernandez2011}, and none incorporate self-tracked personal data in the analytic process.

The following sections describe the interviewer and analyst roles and phases of the \interview framework. Section  \ref{sec:case_study} provides details on how we prepared and performed these interviews through a case study with asthmatic families living with an indoor air quality sensing system.  Section \ref{sec:results} outlines our case study outcomes.  We also provide an online guide~\cite{dataEngagementInterviewGuide} with materials for preparing \interviewsNS, recommendations for selecting an analyst, and a sample protocol for how we structured our own \interviewsNS.

\subsection{The interview team} \label{sec:interview_team}

We performed our \interviews by adding a dedicated data analyst to a pair interviewer team~\cite{monforte2021pairinterivew}. In this arrangement, we divided interviewing responsibilities between two interviewers and left real-time analysis tasks to the analyst. One interviewer was responsible for prompting participants to articulate what they want to do with their data, and the other interviewer kept track of the overall conversational flow. Depending on interviewer experience or subject matter complexity, other research teams may be able to perform \interviews with a single interviewer and analyst.  If performing \interviews using a single interviewer, however, this sole interviewer will be simultaneously responsible for engaging the participant in conversation; tracking various opportunities, comments, or observations worth digging into; and making sure to keep the interview on track and on time. In our \interviewsNS, we found that pair interviewing reduced the cognitive burden on the individual interviewers, and allowed for more focused and productive \interviews in line with the experience of other pair interviewer teams~\cite{monforte2021pairinterivew}.

Our data analyst was responsible for implementing participants' directions for processing their data and communicating the analysis results back to them in an understandable way. We encouraged the analyst to interact with the participants and interviewers to gain any necessary  clarifications or analytic details for completing their analyses,   although we cautioned the analyst not to actively comment on or suggest analysis options so as not to steer participants' choices or behaviors. We prioritized candidates with strong analytic and interpersonal skills when evaluating potential analysts, in order to select someone comfortable with analyzing data on-the-fly during an interview, while also taking direction and communicating with both researchers and participants. Good candidate analysts ideally are: experienced working with data similar to what they will process during the \interviewNS; fluent in their preferred programming language and processing environment; and able to exercise good visualization and interview techniques. 

\subsection{Interview materials}

The interview team brings with them an enriched and formatted collection of the participant's personal data, and a suite of tools and devices for conducting real-time analysis during the \interviewNS. We used Jupyter notebooks for conducting our real-time data processing, but we speculate that other interactive platforms such as Power~BI or Observable can be effective. Although what and how much data to prepare can depend on the broader project aims or goals, it was our experience that participants' questions required access to external data sources to more fully contextualize and support  analysis goals. For our participants, this preparation meant adding local weather and outdoor air quality measures to help further contextualize their own personal indoor air quality data. We benefited from performing pilot interviews, brainstorming, and drawing insights from the literature prior to conducting interviews to help us identify what kinds of additional data sources were likely to be helpful in our context.

Labeling, organizing, formatting, and cleaning data are important steps for any analysis project, and can take an estimated 80\%- 90\% of the effort in data analysis work \cite{kandel2011wrangler,rattenbury2017principles}.  It is important, however, to consider how these transformations may impact or slow down an analyst's ability to handle unanticipated analysis requests during the interview.
Breaking data into separate files can make certain analyses more modular, although the choice to segment data versus maintaining one large data structure can affect its accessibility. For example, in our interviews we experienced that population-wide summaries and between-participant comparisons became more time consuming to compile when this information was separated across different files and directories. These decentralized data caused our analyst to spend time during some interviews reformatting and wrangling several disconnected sources in order to address unanticipated questions, ultimately exposing the limits of the interviews' real-time capabilities.

The tools and techniques used to process personal data may also impact \interviewsNS. We came to our interviews prepared with a variety of read-to-apply analytic techniques based on what we suspected our participants may request and what we knew their data could support, enabling our analyst to quickly perform a variety of common requests in our interviews. Based on our observations of how participants engaged with their data, we also recommend considering the kinds of entry points \cite{kirsh01work} that participants may take into their data. For example, in our interviews, the participants often wanted to jump into their time-series data at a particular season, month, day of week, hour of day, or combination of these conditions. Anticipating entry points that rely on aggregations or data cuts can aid in quickly addressing participants' analysis requests.

Performing the data engagement interview will require that the interview team come prepared with a laptop and external monitor for showing data to a participant.  We also suggest that the analyst prioritize creating visualizations that participants can easily read and understand. For participants comfortable with processing numerical information, standard statistical charts  such as line charts, bar charts, and scatterplots should be sufficient \cite{lee2019correlation, galesic2011graph}. We recommend Munzner's \textit{Visualization Analysis \& Design } \cite{munzner2014visualization} or Ilinsky and Steele's \textit{Designing Data Visualizations} \cite{iliinsky2011designing} as starting points for  researchers interested in learning more about designing effective visualizations.
Standard interview materials such as audio/video recording equipment, note-taking, or sketching supplies are also good practice.

\subsection{Interview phases}

\begin{figure*}
      \centering
      \includegraphics[scale=0.42]{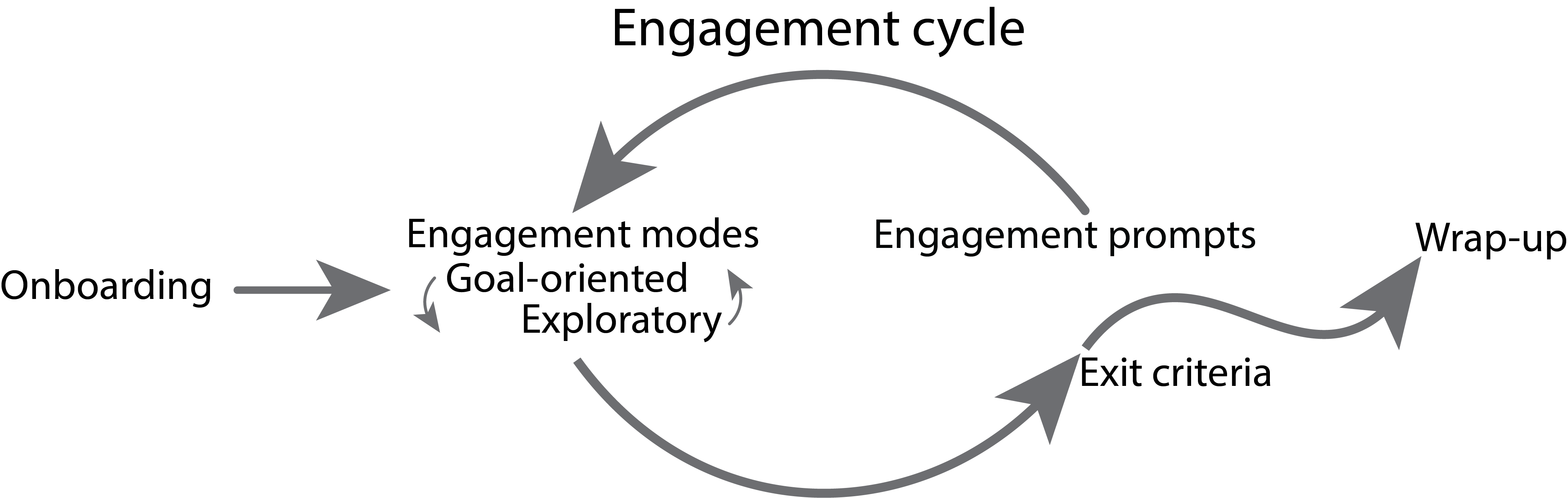}
      \caption{The \interviews began with an introductory stage to remind participants of the goals and scope of the interview.  Incorporating personal data into the interview transitions to an engagement cycle, where interviewers can  guide the participant between engagement activities and research questions. Engagement activities are also cyclic, and can switch between exploratory and goal-oriented modes.  The exploratory engagement mode starts with data, and relies on curiosity or surprise to determine what questions participants will use to direct the analysis process. Goal-oriented engagements use participants' prepared questions or goals for determining how they engage with data and the analysis tasks they undertake. Interviewers can pose various engagement prompts throughout the interview, either in response to participant comments and data engagements or while waiting for analysis results.}~\label{fig:interview_method_cycle}
    \end{figure*}

We divided our \interviews into three distinct phases: onboarding, the engagement cycle, and wrap-up. Both the onboarding and wrap-up phases align with traditional interview practices, whereas the engagement cycle is a unique and critical phase of \interviewsNS. Figure \ref{fig:interview_method_cycle} illustrates an overview of these phases. 

Our \textbf{onboarding phase} introduced the participants to the overall goals of the interview and the format the interview would take. This phase includes introducing and explaining the role of the data analyst, along with the scope and scale of the data they have access to during their interview. During this phase, we actively primed our participants for engaging with their data by having them reflect and expound on their personal data engagement goals, along with prompting them to further operationalize those goals into more specific and concrete analysis tasks. 
Next, the interview enters the \textbf{engagement cycle}. This interview phase cycles between the participants engaging with their data via the data analyst, and the interviewers prompting the participants with questions, comments, or observations that are meant to surface research insights and feedback. The participants can engage with their data using either an exploratory or goal-oriented strategy; these strategies are themselves cyclic, and provide the interviewers some flexibility during the interview for eliciting productive engagements. 

\textbf{Exploratory engagement} is a bottom-up approach occurring serendipitously when the participants explore their data out of curiosity without a concrete goal in mind.
We elicited this type of engagement by showing our participants some part of their data that we thought they might find interesting. Our participants also engaged in an exploratory approach when they inadvertently become distracted while reviewing data for some other goal. Distraction played a prominent role in our interviews, and came from participants' surprise and curiosity when they encountered unexpected features in their data. When participants engaged with their personal data with an exploratory approach, we would use their curiosity and surprise as an opportunity to encourage them to generate a direct question about the data. This prompt often transitioned them to a more goal-oriented engagement.
 
\textbf{Goal-oriented engagement} is a top-down approach occurring when the participants engage with their data in a goal-oriented way by posing a question and directing the analyst to process the data in service of answering that question.
We found that this mode of engagement pushed our participants to grapple with how to both operationalize their questions and interpret the results. During goal-oriented engagements, surprises often distracted our participants, and they reverted to an exploratory approach until they identified a new question or we guided them back to their original goal.
    
Once our participants began to productively engage with their data, we posed questions and made comments or observations that served our goals as researchers. These \textbf{engagement prompts} leveraged the participants' engagement with their data and had them answer questions about their analysis goals, preferences, and approaches; motivated the participants to improve compliance with self-tracking activities; or supported the participants in taking what they learned from their data to make positive changes in their lives. We injected an engagement prompt in response to a specific participant action or statement, or to fill space if there was a lull in the interview. These engagement prompts resemble the traditional semi-structured interview prompts for questioning or clarifying participants' statements \cite{mccracken1988long,leech2002asking}. 

Engagement strategies and prompts can feed into one another from the conversational dynamic that arises around engaging data \cite{Arias-Hernandez2011}. The dynamics of our interviews  shifted between engaging a participant with their data and engaging them with a prompt to observe \textit{why} they wanted to engage their data, \textit{what} their priorities or goals were in practice, and \textit{how} this changed through access to flexible and personalized analysis. We ended our engagement cycles when the interviews reached  the time or energy limit of the participant, a satisfying result for the participant, or saturation of insights and goals of the interviewers.

Finally, the interview enters the \textbf{wrap-up phase}. We thanked our participant and summarized the interview trajectory to provide closure, as well as re-state the study goals to explain how this interview fit into our broader research to further validate the participant's efforts.

\section{Applying the interview framework: An illustrative case study with asthmatics} \label{sec:case_study}

We conducted seven \interviews with participants from a longitudinal study on how asthmatic families engaged with personal indoor air quality data. Our initial goal with conducting these interviews was to better understand and identify how our participants might go about analyzing their data to develop design requirements for a future visual analysis tool. The interviews were developed to be completed in 90 minutes, with our participants taking between 50 - 110 minutes (79 minutes average).

\subsection{Recruiting and preparing the data analyst} \label{sec:recruiting-analyst}

We recruited our data analyst from prospective graduate and undergraduate student candidates within our university's computer science, mathematics, and physics departments.  These candidates came from other researchers' direct recommendation, and in response to a \$17/hr work study position for an interactive data analysis project. We briefed applicants on the nature and goals of the interview method and provided test data sets similar in scope and content to participants' data in preparation for a live-coding interview. The interview process involved the analyst using this test data set to work through several sample questions modeled after those participants had asked in their deployments.  Our recruited analyst was a physics graduate student with extensive experience processing large cosmic ray data sets. This background allowed him to easily handle our time series sensor measurements and to repurpose signal processing scripts to help bootstrap filtering and aggregating our air quality data.

The data analyst is a vital component of \interviews and requires both strong analytic and interpersonal skills. Our data analyst refined his analytic toolbox and interview skills through his experience participating in our second-round pilot interviews. In preparation for the primary interviews, they compiled their accumulated analysis scripts into a workbook that allowed him to quickly execute commonly requested data analysis tasks, as well as create and customize data visualizations using minimal commands. This preparation saved him time and helped make real-time data analysis a reality for the \interviewsNS.

\subsection{Collecting and wrangling participants' data}

\begin{figure*}
    \centering
    \includegraphics[scale=0.26]{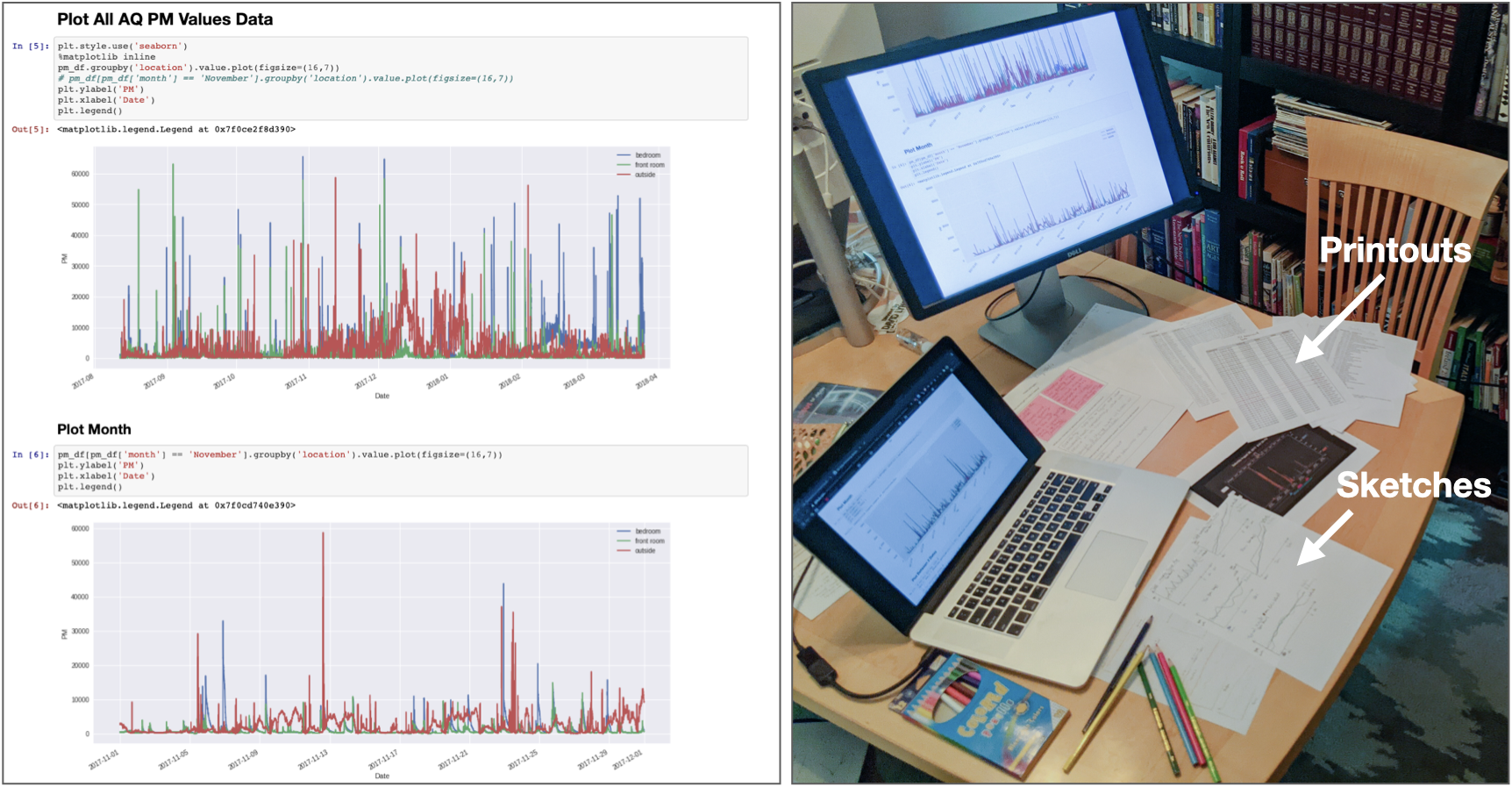}
    \caption{Left: A screenshot of the data analyst's Jupyter notebook used to process and display participants' data.  This view was mirrored between the analyst's laptop and external monitor when presenting data to participants. Right:  A typical \interview setup.  Interviews included physical printouts of a participant's data to help them to understand what data was available. Some participants also used sketching  to communicate how they imagined using their data to the interviewers and data analyst.}
    \label{fig:data_printouts}
\end{figure*}

In preparation for conducting our \interviewsNS, we collected, enriched, and formatted each participant's deployment data, and extracted representative subsets to present as physical printouts during each of their interviews. An example is shown in Figure \ref{fig:data_printouts}. Previous work outlines the data collected through these deployments \cite{Moore2018}. In summary, participants' deployment data consisted of 3 air quality monitors that logged measurements at 60-second intervals over several months; a table of algorithmically detected spikes \cite{smoothedZScore} from  each monitor data stream, including the time, location, monitor ID, and spike value; a table of outbound text message alerts sent to participants based on these detected spikes, including the message timestamp, content, and spike location; and a table of participants' replies to these messages, including the reply timestamp, content, and annotation source (text, tablet, or Google Home). 
    
Based on earlier participant feedback, we further supplemented these data sources with daily EPA Air Quality Index classifications\footnote{https://www.airnow.gov/aqi/}, self-tracked respiratory health surveys collected through a parent medical study \cite{PRISMS}, and environmental data including ambient temperature and humidity, also measured by the air quality monitors. We integrated these additional sources to further contextualize participants' data, and to provide a richer set of analysis opportunities than would be possible from air quality data alone \cite{Bentley2013,tolmie2016has,jones2015exploring}.
    
We formatted these data to support a number of anticipated data cuts \cite{epstein2014taming}, and prepared scripts to filter and facet participants' annotated air quality data according to various questions we had received throughout the study. Examples of these processing scripts and data files are available in our online guide \cite{dataEngagementInterviewGuide}. Basic data processing allowed us to plot filter, group, and aggregate raw air quality measurements by individual participants, sensor locations, or time spans. Further processing also allowed us to cut this information by other temporal characteristics, such as assigned categorical labels like mornings, weekends, seasons, etc., or participants' own annotations that we knew would be relevant to how they thought of their indoor air quality. Other derived data, like the times and locations of detected spikes in participants' air quality data, provided more opportunities to partition and review this information on a spike-by-spike basis.  We additionally coded participants' annotations with representative class labels to provide more categorical filter criteria such as supporting data comparisons between ``cooking'' and ``cleaning'' annotation types. The EPA Air Quality Index and participants' self-tracked asthma surveys provided more options for categorical and contextual processing.  These parameters helped us expand the analysis space for flexibly reviewing participants' data and helped us prepare for a wide range of potential questions during the \interviewsNS.  

\subsection{Other materials}    

We conducted  \interviews in participants' homes, requiring us to come prepared with all necessary interview  data, equipment, and supporting materials. Each interview included a laptop for analysis, an external monitor for sharing analysis results with participants, and an interviewing kit for each participant.  This kit contained a worksheet to capture participants' analysis goals and physical printouts of their self-tracked annotations, detected air quality events, asthma control test scores \cite{jia2013asthma}, and representative subsets of any sensor measurements (Figure \ref{fig:data_printouts}). 

These materials acted as visual aids and physical tokens during the interview to help explain the scope and scale of participants' available personal data in the onboarding phase, and to help them reflect on their data engagement goals within the engagement cycle. We also brought pens and paper to support them with externalizing their analysis process through sketching if they preferred. 

\subsection{Onboarding}
    
In our \interviewsNS, the onboarding phase began ahead of the actual interview. Participants received an e-mail from the interviewers describing the interview’s purpose and our request for them to come prepared with some questions to apply to their data:   \textit{``Imagine you've monitored and logged your home's indoor air quality for the past year.  What would you want to do with it?  What would you want to know?''}   At the start of the interview, we began by revisiting this prompt and having the participant fill out a worksheet to capture their questions, motivations, and goals for engaging their data while we set up our processing environment. This worksheet was used to help focus the participant's thoughts and to serve as a visual reminder of their goals throughout the interview. In 2 of the 7 interviews, participants neglected to prepare ahead of time. In anticipation of this possibility, we had prepared a collection of sample questions and offered them as options to chose from.  
     
After participants completed their worksheet, we had them present their goals and questions, and asked them to explain why they made their choices. Finally, we reminded the participant of their previous year-long indoor air quality deployment \cite{Moore2018} and the data that were collected using physical printouts of representative samples of the participants' own data, such as those shown in Figure \ref{fig:data_printouts}. Using these printouts, we talked through what was available to them during the interview, and what each of these data sources contained.
     
Whereas some participants were ready at the end of the onboarding phase to direct the analyst on how they wanted to load and analyze their data, others were more hesitant or unsure.  In these situations, we performed a short mock exercise between the interviewers and analyst with local weather data to role-play basic analysis tasks and representative participant/analyst interactions.  We performed this exercise in an attempt to lower barriers or anxieties around engaging with the analyst so that participants might more freely take control of the interview once their data were loaded.  

\subsection{Data Engagement}
    
After the onboarding phase, the analyst loaded and presented the participant's data in an overview visualization, with interviewers encouraging the participant to direct the analyst on the ways they wanted to analyze or explore their data. Participants who were ready to dig into one of their prepared questions transitioned to goal-oriented engagement and began to direct the analyst to process data according to how they imagined approaching answering their question. We prompted these participants to explain their thought process and operationalization strategy. These discussions revealed details related to people's analytic approach and assumptions about their data. Other participants used the overview visualization to initiate exploratory engagements. For example, when initially viewing their air quality data, some participants asked to zoom in on prominent air quality spikes. 
    
Outliers or deviations in air quality or health survey data frequently drew participants' attention while they engaged with their data.  These features were a common source of distraction that shifted them from a goal-oriented to an exploratory engagement mode.  Other distractions came when overlaying additional data sources, with many participants attempting to contextualize or correlate trends in one data source to features in another. In particular, several participants were interested to assign causation between their indoor air quality and health outcomes, which shifted their focus away from prior goals, and toward combining data sources to search for potential correlations. When participants became distracted, we prompted them to understand why, what was interesting, and whether they wanted to modify or change their question based on what they saw.  Some participants took the opportunity to continue with a new focus, and others would shift back to their original question.

Throughout our \interviewsNS, we found that many participants quickly cycled between exploratory and goal-oriented engagement facilitated by the dynamic data analysis afforded by the interview framework. When participants were presented with new data, there was an immediate period of review and reflection, during which the interviewers prompted the participants to reflect on their path to answering a question, and to share how what they saw affected the way they  thought. These reflexive prompts led to more and different questions that encouraged the participants to give the data analyst new directives, restarting the engagement cycle once again. 
    
In all but one interview, the interviewers continued eliciting cycles of engagement until the participants reached at least one satisfying answer to one of their questions; in these interviews, the interviewers wrapped up the discussion once they reached the participants' time or energy limit. In the one interview that did not reach a satisfying answer on the part of the participant, the interviewers wrapped up once they reached saturation and were no longer extracting new information from their engagement prompts.

\subsection{Engagement Prompts}

Our engagement prompts throughout the interviews served four purposes. First, we used prompts to encourage and elicit data engagements: \textit{So, with the data we have, is there anything that you would want to see now? What do you want to see next? Does this data match what you thought you would see? What is important to you from this result?} Second, we prompted participants about their analysis strategies to gain insights for the design of a future visual analysis tool: \textit{How confident are you that something like this gets at what you wanted to know? Does this satisfy the question you have? Based on all of the data you've seen, how do you feel it addressed the question you had at the beginning?} Third, we reserved our final engagement prompt to collect observational data on people's operationalization abilities:  \textit{Knowing what you know right now, how would you use your data if you had to go back and do this again?}  Fourth, we prompted participants to give feedback on the \interview method: \textit{How was this process for you? How was interacting with the data analyst? Did anything feel too slow or rushed?}

\section{Interview framework outcomes}\label{sec:results}
    
This section reports on our observed \interview outcomes across seven primary participants.  These outcomes speak to the interview's strengths at engaging and teaching participants through data analysis, teaching researchers through observing participants, and supporting the design and outcomes of personal data studies.

\subsection{What participants say versus what they do}\label{sec:tension}

There is often a (big) difference between what people say they want to do with technology, and what they actually do when given the opportunity \cite{beyer1999contextual}. In almost every \interview we conducted, we observed participants asking unanticipated questions and approaching their analysis in unexpected ways once they began to actively engage with their data.

For example, at the start of P4's interview, she stated her interest to understand how her family's indoor air quality might have affected her daughter's asthma. She said she wanted to overlay periods of data that contained spikes with her daughter's self-tracked asthma data. The analyst began by pulling up an initial overview of her data:
\begin{quote}
    \textbf{Interviewer: }Based on what you wanted to see, how would you like to use this data to answer your question?\\
    \textbf{P4: }Actually, now that I'm looking at [the data], I would also like to see how [spikes in the bedroom] correlate to, for example, vacuuming or something. 
\end{quote}
P4 noticed spikes in the data stream from the sensor in her daughter's bedroom that were much larger than readings from the other sensors. This realization pivoted the interview toward exploring the sources of various spikes and how they correlated (or not) across the various sensors in her deployment. From this comparative analysis task, she learned that certain types of cleaning activities impacted her air quality more than others, which led to a broader conversation on alternative goals and more questions.

Curiosity and surprise also triggered other participants to alter their goals, exposing a possibly more authentic portrayal of what they wanted to do, and could do, with their data. At the start of P3's interview, she told us she wanted to get a sense of whether the air quality in her home was good or bad, and what changes she could make to improve her home's air quality. She initially stated that she wanted to look at large spikes and their annotations to find possible patterns. While exploring her data, however, an especially prominent spike captured P3's attention:
\begin{quote}
    \textbf{P3:} Okay. Wait, go back one more. That one, I want to see that one, because that's weird.
\end{quote}
At this point, P3 switched from an exploratory engagement to one that was goal-oriented in attempt identify the source of this specific  unannotated event.  She went on to identify other interesting spikes that had gone unannotated and realized that answering her questions relied on richly annotated data, prompting frustration that she had not been more diligent in self-tracking events during the deployment.

Participants' interests occasionally even contradicted their own stated goals within the same interview:
\begin{quote}
    \textbf{Interviewer:} Would you ever look back to see when you were sick and look at the air quality then?\\
    \textbf{P5:} No, it would just be for right then, on demand.\\
 \textit{[...Loads data showing previous spikes...]}\\
    \textbf{P5:} Okay, that's pretty interesting. That makes me wonder if it was all through the house [...] That's a question that I have, was that at that same time I was sick? 
\end{quote}
P5's stated interest to review his indoor air quality had always been motivated by wanting to know what was happening around him in the moment \cite{Moore2018}.  
The process of reviewing his data, however, made him engage deeply with exploring the source and potential health impact of an earlier pattern of late-night spikes, going against his claim minutes earlier about his disinterest in retrospective analysis.

Providing feedback on data without actively engaging with it required our participants to \textit{imagine} how their data may support their goals, or which aspects they suspected would be relevant or interesting to review.  Counter to other research techniques that do not incorporate personal data, our observations from conducting \interviews illustrate how directly engaging with personal data can refocus participants' attention, resulting sometimes in new and different goals.

In personal informatics studies, this difference between what people say and do is a critical one when trying to understand how access to data can impact people's lives. Without an understanding of what people \textit{actually} do, researchers and practitioners run the risk of designing tools for the wrong tasks, misunderstanding people's relationship with their data, and missing opportunities to support people in making positive changes in their lives.
 
\subsection{Data engagement interviews can be engaging}\label{sec:engaging}

All seven of our \interviews were successful in getting participants to actively and productively engage with their data. In most cases, participants readily hopped into the engagement cycle and began directing the data analyst. In two interviews, however, the participants required a significant amount of initial prompting by the interviewers to \textit{really} engage with their data.

One example of this reluctance was the interview with P4, who had been much less engaged during her deployment than other participants \cite{Moore2018}, and did not come prepared with a question for her \interviewNS. When discussing her goals with P4a, who made a brief appearance in P4's interview, she described her overall detachment from the process:
    \begin{quote}
        \textbf{P4:} I was not well prepared. You can do a better job. Here's the questions. You can see how much I wrote. [points to blank page]
    \end{quote}
The interviewers were undeterred and discussed aspects of P4's deployment that they knew were important to her.  Despite her unpreparedness, reviewing her data and annotations reminded P4 of the challenges she faced with managing P4a's health. By the end of the interview, P4 was actively engaging with her data to explore what role her indoor air quality may have played in impacting P4a's health. During the wrap-up phase, she apologized for her initial reluctance at the start of the interview and said:
    \begin{quote}
        \textbf{P4:} I know I wasn't really well prepared. [...] I kind of forgot our ultimate reasons for doing it. But looking over these things, I remember now what we were dealing with two years ago. Like I said, I had kind of forgotten about that, but looking at this, I do remember the struggles we were having and what we were trying to do to figure things out. And having this information back then probably would've been very helpful.
    \end{quote}
P4's distance from the initial deployment made her forget many of her goals for participating, but her ability to directly engage with her data surfaced several memories and quickly led to multiple questions, ultimately providing her with valuable insights. P4 even quipped at the end of her interview that she had lost track of which question we were working on from having posed so many. 

The \interviews were also successful in engaging new members of the participating families. During the deployment stage (S1), both the interactions with the air quality monitoring systems and the interviews were predominately led by a single, primary caretaker in the home \cite{Moore2018}. Postdeployment interviews confirmed these disengagements:
    \begin{quote}
        \textbf{Interviewer:} Could you describe your level of involvement with the [deployment]?\\
        \textbf{P1-S:} Minimally.  If the app crashes, I’ll turn it back on.  If I see a big spike, I’ll ask [P1] about it.  But I’m not doing anything differently because of it.
    \end{quote}
Family members like P1-S attributed their indifference to not being personally affected by air quality, not having the interest or time to engage with their air quality system, or the distribution of labor within the home.    

During the \interview with P1, however, P1-S entered the room and stopped to look at the data:
    \begin{quote}
        \textbf{P1-S}: So it looks like inside gets worse than outside.\\
        \textbf{P1:} It's hard to tell...\\
        \textbf{P1-S:} But everything spiked.
    \end{quote}
P1-S proceeded to pull up a chair and take part in the rest of the interview, directing the analyst to review indoor air quality spikes as a way of comparing their indoor and outdoor air quality.
A similar engagement happened with P2's husband during her \interviewNS. 

The involvement of P1-S and P2-S in the \interviews surprised us, as we had made considerable efforts to engage multiple family members during the initial system deployments and again in postdeployment interviews, but with limited success \cite{Moore2018}. Making the families' personal data accessible and analyzable through this interview framework succeeded in generating more interest and broader family engagement than our previous traditional deployment and interview practices. Effecting this level of engagement with participants and previously disengaged family members lends further evidence of \interviewsNS' promise, both as a research method and an outreach tool for generating interest with personal data.

\subsection{Engagements can teach participants new things}\label{sec:learning}

During \interviewsNS, many of the participants learned new things about their home's air quality, acquired new data analysis insights, and became more confident in their analysis skills. For example, while reviewing data during the interview, P1-S~---~who had not engaged with the data during the initial deployment~---~voiced a strong opinion that his home's indoor air quality was quite poor. P1, looking at the same data, drew different conclusions. This disconnect between P1 and P1-S's interpretations led to a broader discussion between the two of them on how the deployed sensors measured air quality data, and how to interpret those measurements. P1-S directed the analyst throughout the conversation to process and display the data in a variety of ways, which P1 used to explain differences between how personal activities affect their indoor air quality, compared to how outside air quality conditions influence the air quality inside their home. Eventually, P1-S and P1 arrived at a shared understanding, backed up by the data: 
    \begin{quote}
        \textbf{P1-S: } It looks like ... Maybe I was wrong, okay. I was just curious, because we hear that all the time ... They're just like, ``Stay indoors. It's bad outside.'' But then I've seen way more spikes indoors than I would've guessed. I'm just curious if, even if you're indoors with all the windows closed, do you still see a spike indoors on the bad [outdoor] days?
    \end{quote}
    In this situation, the ability to jointly engage with data resulted in a productive conversation between participants, resulting in P1-S broadening his understanding of air quality, revising his position, and eventually articulating a more actionable question to explore.

    Working closely with data led other participants to recognize analysis limitations when trying to correlate data  collected over different timescales. P2-S directed the analyst to overlay air quality measurements from real-time sensors to P2's  self-tracked and weekly aggregated asthma survey scores:
    \begin{quote}
        \textbf{P2-S:} Maybe you could see trailing [asthma results]. Like if there were high [air quality spikes] on Monday, maybe Tuesday your [asthma] values would just suck [...but these surveys are] weekly, so you couldn't really tell.
    \end{quote}
    P3 describes a similar limitation:
    \begin{quote}
        \textbf{P3:} If he has an asthma attack and there was a spike in the kitchen, if I'm using weekly [asthma] data, I'm not sure I can correlate the spike with his asthma, because maybe it was when he was camping, but I might not know when I come back and look at the information.
    \end{quote}
    Whereas P2-S identifies limitations of trying to attribute specific real-time events with a highly aggregated score, P3 identifies the problem of correlating two different data sources without contextual information that would indicate whether these sources do indeed have a dependency. Both of these realizations came from directly engaging with the data and were unprompted from the interviewers. Our previous participant interviews in S1 had recorded their frequent requests to compare and correlate air quality data to health data, yet it was not until participants directly attempted this comparison that they realized inherent challenges that come from working with diverse and independently collected data.

    Other participants learned that data analysis was not as daunting as they assumed it would be.  Given P4's general lack of engagement throughout the deployment stage of our longitudinal study, we were surprised to find her engaging and directing the analyst by the end of her \interviewNS.  In spite of her initial lack of confidence or interest for engaging her data, she felt more comfortable in her ability to engage and analyze her own data by the end:
    \begin{quote}
        \textbf{Interviewer:} Seeing the data in this way, how confident do you feel being able to draw some of these conclusions?\\
        \textbf{P4:} I think I could probably do it. For me it would be more a matter of how simple the program is. I am very bad with computers, so it would have to be very simple. But maybe just talking about it, just using this information and applying it, I could do that. It doesn't seem hard.
    \end{quote}
   Despite P4's initial lack of confidence, working with an analyst helped her to develop a sense of competence that she may not have acquired through use of a custom system.  We attribute this change of behavior to lowering the analysis barrier, and allowing P4 to focus on her data and what it could show her, rather than the analytic steps or perceived skill required to show it.  In fact, all  participants appreciated working directly with an analyst, and often remarked how this collaboration helped make the process, and their data, more understandable:
        \begin{quote}
            \textbf{P2:}  I think because we were working together it wasn't confusing, and you told me what was going to change. [...] it was helpful that you asked me to organize it, otherwise I would be like... this doesn't matter to me, why would I look at this, you know?  
        \end{quote}
        
        \begin{quote}
            \textbf{P4a:} If you just showed me all this stuff, I'd be very confused. But with [the analyst], I think it helps people understand it better [...] I think it's a very flexible way for you guys to cooperate with the participants, and say like, ``Is that what you're asking about?'' And then answering the questions about that.
        \end{quote}
        
        \begin{quote}
            \textbf{P5:} Overall, being able to see [my data], especially interacting with the questions, was kind of cool. I thought it was kind of cool to be able to actually have a question, and look at [the data]. [The analyst] changed four words, and three numbers, and boom, boom, boom, and there [is the answer]. That was pretty awesome, actually.
        \end{quote}
These observations lend evidence that collaborations with a dedicated analyst can help participants to critically engage with their personal data in \interviewsNS, and that these engagements can benefit participants' sensemaking, problem solving, and overall sense of agency.  These benefits also advance our own research goal to observe how to help people gain insights from their personal data.

\subsection{Engagements can teach researchers new things}\label{sec:future-design}

As a team of visualization and personal informatics researchers, we conducted our \interviews with the research intent of acquiring design requirements for a future visual analysis system that could support people in analyzing their personal data. Analyzing these interviews revealed a wealth of new insights into the challenges and opportunities for visualization design. We report in detail on those insights in a companion paper\cite{moore2021gap}; here, we briefly summarize a few of our findings and their implications for visualization design to illustrate the usefulness of \interviews for gaining research insights. We encourage readers to see our companion paper for more detailed validation of the utility of the \interview method.

Given our study context of asthmatic families monitoring their indoor air quality, we approached our participants' \interviews in a serious and goal-oriented way, expecting they would as well.  Once our participants began engaging with their data, however, we observed that they became playful with how they recounted their deployments, and were quick to joke about previous annotations, many around cooking and burning food:
\begin{quote}
    \textbf{P1:} It's funny, I kind of did start putting in snarky remarks in some of the comments, you probably noticed. ‘Cause there was a while that my oven had burnt pizza on the bottom and every time we would turn the oven on [the system] was like “HEY!! HEY!! HEY!!” and, yep, still haven't cleaned my oven. You wanna come clean my oven? 'Cause I still haven't cleaned my oven. [laughs]
\end{quote}
\begin{quote}
    \textbf{P3:} We did the same thing over and over again.  Bacon. [laughs]  I think that like 90\% of our annotations are probably bacon.  I like bacon! [laughs]
\end{quote}
During the interview, P3's playful interest in bacon transitioned into a broader exploration of her cooking habits. This exploration provided P3 with a more holistic view of what was causing poor indoor air quality in her home:
\begin{quote}
    \textbf{P3:} I remember making the connection between the olive oil and the [spikes]. And I also knew that it was kind of every time we cooked bacon there was a [spike]. But I guess I didn't realize how many of them, overall, were actually cooking episodes... Like ``cooking pancakes'', ``cooking eggs'', ``[my daughter] burning the tortillas''. It's all cooking.
\end{quote}

P3 and other participants used play as a mechanism to dive more deeply into their data, which frequently led them to serendipitous discoveries. This type of playful and serendipitous engagement is understudied and undervalued in visualization research, perhaps due to a decades-long framing of visualization as a vehicle for cognitive amplification and insight generation \cite{card1999readings}. Instead, most visual analysis tools are designed for goal-oriented behaviors\cite{shneiderman2004designing,sedlmair2012design,brehmer2013multi}. Our findings suggest that prioritizing play and serendipity in the design of new systems could lead to innovative ways to support people in engaging with personal data. 

As we discussed in Section \ref{sec:learning}, directly engaging with personal data during \interviews helped all our participants learn new things, and in some cases increased their confidence for doing so. Yet, when we asked them directly at the end of the interviews how likely they would be to analyze their data on their own, all but P4a -- our youngest participant -- were reluctant to do so.  Asthmatic parents, and parents of asthmatic children, live lives full of responsibilities.  We witnessed this during our interview with P1, whose parenting and household routines left little time for exploring or analyzing her personal data:

\begin{quote}
    \textbf{P1:} As a busy mom with small children, having to just... [video game noises in the background]
    
    \textbf{Interviewer:} You're busy?
    
    \textbf{P1:} Yeah, busy!  Obviously!  With herding small people [laughs].  For me... it's interesting... I just don't have the time to sit down and look through all the numbers, and do all that stuff.
\end{quote}
For P2, her reluctance stemmed from a concern about medical implications:
\begin{quote}
    \textbf{P2:} I don't know that I would ever just pull it up and look at it for data's sake, if that makes sense?  I'm not a numbers person, I'm not a computers person. If I can look something up and say, "hey, I see this pattern", I can take it back to my doctor and talk about that there, and maybe that helps change a treatment plan, or whatever ... I could see myself doing that.
\end{quote}
In this case, P2's hesitation came from the potential health risk of doing something wrong, and preferring instead to have her doctor interpret her data, rather than risk drawing those conclusions herself.  

We were surprised by our participants' lack of enthusiasm for a visual analysis tool that would enable them to perform the same types of analysis they had engaged with during their interviews given the productive and positive outcomes of those interviews. Furthermore, if we were unable to motivate tool-usage by asthmatics living in an area that frequently experiences some of the worst air quality in the country~\cite{slcair-country}, we wondered how hard it might be to motivate other people living in less dire circumstances. The success of the \interviews with our participants, however, points to opportunities to focus visualization research efforts on designing collaborative social systems rather than just tools.

Our companion paper \cite{moore2021gap} expands on these two ideas~--~designing for play, and designing social systems~--~and presents more results from analyzing our \interviewsNS. The interviews provided us with new insights into how we might build future visual analysis tools and systems that help people to engage with their data. We provide the summary here as evidence for the efficacy of \interviews as a research method.

\subsection{Seeing the value of self-tracking}\label{sec:self-tracking}
    
Like many personal informatics domains, our specific context --- asthmatic families living with an indoor air quality monitoring system --- includes self-tracked data. These data include participant-provided annotations of household activities, such as vacuuming and cooking, and a daily survey that tracked the respiratory health of the asthmatic family members. Although most of our participants maintained a high degree of self-tracking compliance throughout their deployments \cite{Moore2018}, several of them still found the quality of their data lacking when they tried to make use of it during the interview:
    \begin{quote}
         \textbf{P3:} It kind of frustrates me. It would make me want to go back now, knowing that I could have it all, I would be more vigilant about [annotating]. Because I kind of get lax about it. Then I'd be [annotating more] so that I could use that to cross-reference stuff. 
    \end{quote}    
In this case, P3 realized how the lack of content in some of her annotations failed to support the kinds of correlations she was looking to make with her health data.  
    
P6 experienced a similar issue when attempting to use his health survey data to determine whether his home's air quality affected his asthma, but was unable to do so because of his irregular and inaccurate survey responses:
    \begin{quote}
        \textbf{P6:} You know a lot of [the value] is dependent on good sensors and then good data I'm inputting. Knowing how it can be done is going to motivate me to pay more attention to those [survey questions]. Because moving forward, if this is an opportunity to get my raw data, I'm not going to want to see just a bunch of fives. Like there's no way in January, December, I was that fine every week. That to me is just nothing but laziness on my part... Knowing that the information is retrievable motivates me to want to provide more accurate data.  
    \end{quote}
    Through engaging their data during the interview, P3 and P6 came to understand that their self-tracked data are valuable only when they commit to tracking regularly, accurately, and richly. Even the least compliant self-tracker among our participants came to understand the value of her data during the \interviewNS:
    \begin{quote}
        \textbf{P4:} I remember we got a little tired of all the [annotating] and now I feel bad I didn't respond more because I can see how you use this.
    \end{quote}
    
Although we performed these interviews after the participants' deployments had ended, and thus cannot say that the motivation they exhibited during the interview would translate to better self-tracking, we speculate that conducting \interviews early in a study could motivate participants to better comply with self-tracking.  Motivating and maintaining self-tracking is a balance of lowering barriers to reduce capture burden~\cite{cordeiro2015barriers, epstein2015lived} or tracking fatigue \cite{choe2014understanding} while maintaining enough engagement  to not lose a sense of responsibility to the tracking process \cite{Li2012}.  Motivation is directly related to people's willingness to track \cite{Choe2017semiautomated}, making \interviews a useful technique for engaging participants to improve or maintain their self-tracking habit once they see how useful their data can be.

\subsection{Improving field deployments} \label{sec:field-deployment}

Our field deployment design included an alert system that sent a text message to our primary participants when the air quality monitoring system detected a spike in their indoor air quality. We developed this alert system primarily as a mechanism to encourage annotation --- the participants could respond to an alert-text with a short message about any potentially correlated activities occurring in their home. As the alerts were meant to elicit a response by the participants, we created rules that would turn off alerts during nighttime hours.  
    
Ahead of the \interview with P5, we knew that he had been using the air quality data to hold his family members accountable for impacting their home's indoor environment \cite{Moore2018}. We learned during the interview that he had greater asthma symptoms at night, which motivated him to direct the data analyst to look at air quality sensor data and annotations during nighttime hours. He found, however, that nighttime annotations were missing: 
    \begin{quote}
        \textbf{P5:} Well, if we [had the data], I could say okay, who was out in the kitchen cooking because [my son] likes to cook late night snacks, sometimes. But, I guess  we won't be able to get there.
    \end{quote}
    
Although it seemed reasonable to enact text alert rules based on our assumptions of normative family dynamics at the time of designing our field deployment, our assumptions denied P5 opportunities to sufficiently self-track. Worse still, our assumptions and decisions caused P5 to develop an incomplete awareness of his indoor air quality by not alerting him to many nighttime air quality spikes, which he was surprised to observe during the \interviewNS.  Had we performed this \interview early in the deployment, we could have modified our texting rules to better accommodate P5's family dynamics, and enabled him to be aware and make an effort to investigate their cause. We speculate that other field deployments would similarly benefit from using \interviews early in a deployment to challenge normative assumptions built into deployed technology.

\section{Discussion}\label{sec:discussion}

Our case study outcomes lend evidence that \interviews can: engage people with their personal data in analytic contexts (Section \ref{sec:engaging}); teach participants new things through their engagements (Section \ref{sec:learning}); motivate the value of self-tracking by showing people how they can use their data (Section \ref{sec:self-tracking}); and improve our own understanding of how to better design for users' needs (Sections \ref{sec:field-deployment}, \ref{sec:future-design}).  Where our previous high-level participatory workshop feedback was related to things participants \textit{imagined} wanting to do, our choice to incorporate personal data as a core design element in our interviews helped us to directly observe what our participants \textit{actually did} with their data (Section \ref{sec:tension}). We speculate that these outcomes more broadly position this interview method as a viable interview technique for personal informatics and visual analytics research.

\subsection{Flexible, to a point}

During our \interviewsNS, we strove to create the illusion that any analysis was possible, that it could be done in real-time, and that it was made to order for our participants. In practice, however, data analysis is often not an instant-answer kind of endeavor~\cite{kandel2011wrangler,dasu2003exploratory}. The analysis tasks within our \interviews were often nearly real-time due to our data preprocessing and prepared scripts based on the kinds of questions we anticipated our participants might ask. In spite of this preparation, participants still managed to pose questions that surprised us. In some ways this was a victory --- despite all of our previous efforts to find out what they wanted from their data, \interviews still yielded new kinds of questions. Yet, some of these questions also slowed our analysis down to a crawl and shattered the real-time data analysis illusion.
    
When these situations arose in our sessions~---~as they did with P2 P3, P4a, and P5~---~we did not have a predefined plan for how to respond. These breakdowns emerged when attempting analysis tasks that required additional or unexpected data processing, such as aggregation in ways that underlying data organization made difficult, or attempting to answer questions that sat at the outer reaches of what was possible with the participants' available data.  We speculate that these challenges stem from a need to reformat data on the fly, or a lack of access to relevant data in the limited time available for participants' interviews. When encountering these lulls, interviewers filled this downtime by leaning on interview engagement prompts to have the participants anticipate or reflect on what the answer \textit{could} be, while the analyst wrangled with the data. In some of these cases, however, it became obvious that we were stalling, and some participants even apologized for asking the analyst to do something ``hard.'' Based on our experiences, we encourage interviewers to have a plan for what to do when the data analysis requires time. One suggestion is to cut short certain analytic pathways or reject the question outright if interviewers suspect they will not be productive. If a participant is a long-term participant,  another approach is to work on the analysis postinterview and bring it to the participant later.  Having a plan ready for these circumstances can help reduce the chance of taking the interview participant out of a collaborative or analytic head space, or disrupting the \interview flow.

\subsection{Two interviewers are (probably) better}

Our decision to include two interviewers on the interview team was rooted in our concerns about the complexities of \interviews from the interviewer perspective. We anticipated that a single interviewer may find it difficult to remain deeply engaged in both the data analysis process and guidance of a participant, as well as keeping track of the larger interview direction. Adding a second interviewer to our interview team eased the interviewer burdens in our \interviews and  allowed one interviewer to lead the discussion and maintain engagement with the participant, with the second interviewer ensuring the goals of the interview were met and calling attention to anything interesting or surprising the first interviewer may have missed. Our pair of interviewers were also able to discuss their insights and reflections with each other postinterview, and did so as they drove back to their lab from each participants' home. We speculate this pair interviewer approach has the potential to not only increase the quality of findings, but also improve rapport with participants \cite{monforte2021pairinterivew}.

In some circumstances, however, a second interviewer may be considered a liability. Examples of this include instances where multiple interviewers may intimidate some participants, such as when engaging sensitive populations or research topics.  Thus, the decision to utilize two interviewers hinges on whether these potential interpersonal effects outweigh the benefits of sharing interview tasks. For \interviewsNS, we suspect that the complexity of the interview protocol necessitates a pair interviewer, but that the interview team should consider ways to reduce negative effects such as intentionally diversifying the team.

\subsection{Empowerment}

Prior to conducting \interviewsNS, we had struggled to engage other family members throughout our data collection and deployment phases.  We report in previous work that this lack of engagement was driven by a division of labor within the home and a general lack of interest from nonasthmatic family members \cite{Moore2018}. Using \interviewsNS, however, we were able to motivate previously disengaged family members (P1-S, P2-S) to participate in our study. Based on these outcomes, we speculate that it may be possible to advantageously use \interviews for purposefully motivating disengaged participants or family members.  

Although we were able to use these interviews to motivate and engage more people than at previous times in our longitudinal study, not everyone was equally empowered. Many of our female participants expressed low confidence in their participation (P1, P2, P4, P4a), whereas male perspectives were more confident and outspoken (P1-S, P2-S, P5, P6).  We see this shift as potentially rooted in traditional gender role stereotypes associating a feminine identity with caregiving and a masculine identity with math, data analysis, and computing. Similar dynamics exist in the context of smart homes, where incorporating a technological component into household maintenance shifts domestic attitudes toward  ``default to the `household expert'{},'' who is typically male \cite{kennedy2015digital}.  \Interviews with P1 and P2 reflected these dynamics by soliciting, or complying with, their husbands' priorities. This deference stands to alter the interview dynamics and may have obscured P1 and P2's perspectives. Being aware of these interpersonal dynamics can help researchers plan their own \interviews and purposefully guard against collecting or propagating normative views in the design of new data analysis tools.  In this vein, fielding a more diverse interview team may help researchers motivate and engage with traditionally disempowered perspectives, and offers another advantage for having multiple interviewers for this method.

\subsection{Transferability}

We conducted \interviews  with a small number of participants working with indoor air quality data, but speculate that this method can transfer more broadly to other personal informatics domains. This interview framework is not tied to any one kind of personal data, and so this approach can be tailored for various research objectives related to how people review and learn from their data. \Interviews allow participants to apply their own lived experiences to identify, direct, and prioritize analytic tasks for their own particular needs.  Free from this responsibility, researchers can focus on using the engagement cycle to strategically prompt participants and capture relevant research insights.   We speculate this method can be applied in design contexts to capture more actionable and accurate feedback early in research design studies; for helping everyday people to better understand and interpret real-world data processing and interpretation by work through basic analysis tasks using their data; or as a hands-on educational tool for motivating self-trackers to improve or persist in their tracking regimen. 

\Interviews can excel at uncovering how people engage with and analyze data if they have some knowledge about the context of that data, regardless of their backgrounds or analysis expertise. This strength especially lends itself to personal informatics research, where people are experts on their own lives by definition, but often do not have analysis expertise. For the \interview to work well, the data sets need to be rich enough that they lend themselves to analysis tasks, and the data should either be quantitative or easily quantifiable. This approach is less likely to be useful in a situation where the participant has limited or no familiarity with the context of the data, where they genuinely do not care about the data, or where the data do not lend itself to quantitative analysis. Furthermore, some quantitative data sets will be too dense or complex for the analyst to work with in real time while the researcher keeps the participant engaged. For example, accelerometer measurements may be difficult to engage in this context, but if processed into step counts, these data are likely to be engaging. Answering research questions about whether or how participants would engage with the data on their own may also be difficult given the context of the \interviews being so different from the circumstances they would encounter if they were engaging with their data alone.

\Interviews are a new and different tool in the toolbox of HCI methods, including those typically used in personal informatics. The most common method for personal informatics research to this point has been more traditional semi-structured interviews \cite{epstein2020}, occasionally also presenting collected personal data. With interviews, researchers and participants are constrained by the representations and analysis that are available, thus limiting the ability to engage with the data interactively in real time. Participatory design methods are another approach to eliciting needs. These methods share an aspect of interactive engagement, where the participant is an expert. However, they also tend to focus on the goal that there is a tool being designed, rather that the immediate task of engaging the data. In contrast, one takeaway from our results was that perhaps designing a tool is not the best approach for this specific user context. It is difficult to imagine arriving at a similar conclusion with a participatory design framing. 

\Interviews share some similarities with the think aloud method. Both ask participants to provide step by step explanations of their thought process for accomplishing a task --- in our case exploring their data set. However, think alouds involve a participant using an interface without external support or intervention while a researcher observes. \Interviews are distinct from this approach, which require that participants interact with the analyst and the interviewer, rather than solely interacting with an interface. The standard Wizard of Oz \cite{dahlback1993wizard} approach is another similar method that presents an interface to a participant, which is actually powered by a human ``behind the curtain.'' In the case of \interviewsNS, no interface has been designed; the analyst \textit{is} the interface, and no clear reason exists to hide them away. \Interviews draw heavily on the success and value of pair analytics \cite{Arias-Hernandez2011}, but with the key difference that pair analytics requires both the domain expert and the analyst to have similar levels of analytic and computational expertise in order to productively work through a task. The addition of \interviews to the HCI toolbox enables research and data collection with greater flexibility than adjacent methods by eliminating the need for a participant-facing interface or for participant analytic capabilities.

\section{Limitations}\label{sec:limitations}

Although we argue that \interviews can afford a more authentic view of people's personal data engagements across a range of contexts, we acknowledge the potential limitations to the ecological validity of our observations. We speculate that participants' \interviews will have lasting effects for how they think of their data and behaviors, yet more work is needed to explore how \interviews might influence long-term behavior change. Furthermore, despite showing evidence that these interviews advanced our broader research goals to help participants learn more about their data,  the observations of this work are limited to the specific contexts and circumstances from conducting seven \interviews with a small and specialized user-group. Further validation of this interview framework will require broader application in other situations, and comparison to alternative research methods.  We encourage others to use \interviews and explore how this method can be applied in other personal informatics use cases.

Whereas we advocate for a 3-person team of 2 interviewers and 1 data analyst to improve the efficacy of \interviewsNS, we recognize that this team size may pose challenges from an overemphasized power imbalance through disparities in gender, race, numeracy, or socioeconomic status. Examples in this work include our all-male interview team interviewing female participants on their own, or our participants with limited data analysis experiences directing a graduate-educated analyst. These circumstances may have caused some participants to feel less willing to openly discuss their thought processes or analysis ideas, potentially preventing our observational data from accurately reflecting how people might engage with data on their own, or with one another, outside our study context. We encourage other research teams conducting \interviews to be mindful of the trade-off between interview efficacy and power dynamics, and to consider ways to diversify the interview team as a mechanism to reduce possible imbalances.

Finally, conducting \interviews benefit from having a trained data analyst. This suggestion can pose challenges for conducting this method in more remote or underdeveloped locations where it may be difficult to find a suitable candidates to fill this role. As a result, this approach may not scale as broadly as other interview methods or deployed analysis tools. Further research could explore whether remote collaborations are a suitable replacement for an in-person analyst.

\section{Conclusion}\label{sec:conclusion}

This work presents the \interviewNS: an interview method that supports people to deeply engage with personal data. The \interview strikes a balance between the flexible, lightweight user engagement approaches that do not incorporate personal data, and the more custom, heavyweight analytic tools requiring significant design overhead. We outline a general framework for conducting these interviews, and present a case study from performing seven \interviews with our study participants. We speculate that  \interviews can be extended beyond working with personal air quality data, and be applicable across many different personal informatics domains.  For future work, we are interested in conducting \interviews in other contexts, and to continue mining our rich interview results for designing future tools and techniques that support people in analyzing indoor air quality data.

\vspace{-2mm}
\section*{Acknowledgements}
\vspace{-2mm}

The authors thank Greg Furlich for his indispensable data analysis skills, our study participants for sticking with us throughout this longitudinal study, and our anonymous reviewers for their thorough and supportive comments. This research was supported by the National Institute of Biomedical Imaging and Bioengineering of the National Institutes of Health under award number U54EB021973; the National Science Foundation under award number 1936071; and by the Wallenberg AI, Autonomous Systems and Software Program (WASP) and the Knut and Alice Wallenberg Foundation. The content is solely the responsibility of the authors and does not necessarily represent the official views of these funding agencies.

\bibliographystyle{unsrt}
\bibliography{references}

\begin{thebibliography}{10}

\bibitem{vanKollenburg2018Exploring}
Janne van Kollenburg, Sander Bogers, Heleen Rutjes, Eva Deckers, Joep Frens,
  and Caroline Hummels.
\newblock Exploring the value of parent tracked baby data in interactions with
  healthcare professionals: A data-enabled design exploration.
\newblock In {\em Proceedings of the 2018 CHI Conference on Human Factors in
  Computing Systems}, page 1–12, New York, NY, USA, 2018. Association for
  Computing Machinery.

\bibitem{schroeder2019examining}
Jessica Schroeder, Ravi Karkar, Natalia Murinova, James Fogarty, and Sean~A
  Munson.
\newblock Examining opportunities for goal-directed self-tracking to support
  chronic condition management.
\newblock {\em Proceedings of the ACM on Interactive, Mobile, Wearable and
  Ubiquitous Technologies}, 3(4):1--26, 2019.

\bibitem{Tang2017Harnessing}
Lie~Ming Tang and Judy Kay.
\newblock Harnessing long term physical activity data—how long-term trackers
  use data and how an adherence-based interface supports new insights.
\newblock {\em Proc. ACM Interact. Mob. Wearable Ubiquitous Technol.}, 1(2),
  June 2017.

\bibitem{choe2014understanding}
Eun~Kyoung Choe, Nicole~B Lee, Bongshin Lee, Wanda Pratt, and Julie~A Kientz.
\newblock Understanding quantified-selfers' practices in collecting and
  exploring personal data.
\newblock In {\em Proceedings of the SIGCHI Conference on Human Factors in
  Computing Systems}, pages 1143--1152. ACM, 2014.

\bibitem{Marasoiu2016}
Mariana Marasoiu, Alan~F. Blackwell, Advait Sarkar, and Martin Spott.
\newblock {Clarifying hypotheses by sketching data}.
\newblock {\em Proceedings of EG/VGTC Conference on Visualization (EuroVis
  2016)}, 2016.

\bibitem{Thudt2018}
Alice Thudt, Uta Hinrichs, Samuel Huron, and Sheelagh Carpendale.
\newblock {Self-Reflection and Personal Physicalization Construction}.
\newblock {\em Proceedings of the 2018 CHI Conference on Human Factors in
  Computing Systems - CHI '18}, pages 1--13, 2018.

\bibitem{tolmie2016has}
Peter Tolmie, Andy Crabtree, Tom Rodden, James Colley, and Ewa Luger.
\newblock “this has to be the cats”: Personal data legibility in networked
  sensing systems.
\newblock In {\em Proceedings of the 19th ACM Conference on Computer-Supported
  Cooperative Work \& Social Computing}, pages 491--502. ACM, 2016.

\bibitem{epstein2014taming}
Daniel Epstein, Felicia Cordeiro, Elizabeth Bales, James Fogarty, and Sean
  Munson.
\newblock Taming data complexity in lifelogs: exploring visual cuts of personal
  informatics data.
\newblock In {\em Proceedings of the 2014 conference on Designing interactive
  systems}, pages 667--676. ACM, 2014.

\bibitem{Moore2018}
Jimmy Moore, Pascal Goffin, Miriah Meyer, Philip Lundrigan, Neal Patwari,
  Katherine Sward, and Jason Wiese.
\newblock Managing in-home environments through sensing, annotating, and
  visualizing air quality data.
\newblock {\em Proceedings of the ACM on Interactive, Mobile, Wearable and
  Ubiquitous Technologies (IMWUT)(Ubicomp '18)}, 2(3), Sept 2018.

\bibitem{dataEngagementInterviewGuide}
Jimmy Moore.
\newblock The data engagement interview guide.
\newblock \url{https://vdl.sci.utah.edu/EngagementInterviews/}, 2021.
\newblock Accessed: 2021-10-15.

\bibitem{moore2021gap}
Jimmy Moore, Pascal Goffin, Jason Wiese, and Miriah Meyer.
\newblock Exploring the personal informatics analysis gap: ``there's a lot of
  bacon``.
\newblock {\em In submission IEEE VIS 2021}, x(x), September 2021.

\bibitem{kay2012lullaby}
Matthew Kay, Eun~Kyoung Choe, Jesse Shepherd, Benjamin Greenstein, Nathaniel
  Watson, Sunny Consolvo, and Julie~A Kientz.
\newblock Lullaby: a capture \& access system for understanding the sleep
  environment.
\newblock In {\em Proceedings of the 2012 ACM conference on ubiquitous
  computing}, pages 226--234, 2012.

\bibitem{consolvo2006design}
Sunny Consolvo, Katherine Everitt, Ian Smith, and James~A Landay.
\newblock Design requirements for technologies that encourage physical
  activity.
\newblock In {\em Proceedings of the SIGCHI conference on Human Factors in
  computing systems}, pages 457--466, 2006.

\bibitem{consolvo2008activity}
Sunny Consolvo, David~W McDonald, Tammy Toscos, Mike~Y Chen, Jon Froehlich,
  Beverly Harrison, Predrag Klasnja, Anthony LaMarca, Louis LeGrand, Ryan
  Libby, et~al.
\newblock Activity sensing in the wild: a field trial of ubifit garden.
\newblock In {\em Proceedings of the SIGCHI conference on human factors in
  computing systems}, pages 1797--1806, 2008.

\bibitem{epstein2016beyond}
Daniel~A Epstein, Monica Caraway, Chuck Johnston, An~Ping, James Fogarty, and
  Sean~A Munson.
\newblock Beyond abandonment to next steps: understanding and designing for
  life after personal informatics tool use.
\newblock In {\em Proceedings of the 2016 CHI Conference on Human Factors in
  Computing Systems}, pages 1109--1113, 2016.

\bibitem{kim2009inair}
Sunyoung Kim and Eric Paulos.
\newblock inair: measuring and visualizing indoor air quality.
\newblock In {\em Proceedings of the 11th international conference on
  Ubiquitous computing}, pages 81--84, 2009.

\bibitem{kim2010inair}
Sunyoung Kim and Eric Paulos.
\newblock Inair: sharing indoor air quality measurements and visualizations.
\newblock In {\em Proceedings of the SIGCHI Conference on Human Factors in
  Computing Systems}, pages 1861--1870, 2010.

\bibitem{kim2013inair}
Sunyoung Kim, Eric Paulos, and Jennifer Mankoff.
\newblock inair: a longitudinal study of indoor air quality measurements and
  visualizations.
\newblock In {\em Proceedings of the SIGCHI Conference on Human Factors in
  Computing Systems}, pages 2745--2754, 2013.

\bibitem{fang2016airsense}
Biyi Fang, Qiumin Xu, Taiwoo Park, and Mi~Zhang.
\newblock Airsense: an intelligent home-based sensing system for indoor air
  quality analytics.
\newblock In {\em Proceedings of the 2016 ACM International joint conference on
  pervasive and ubiquitous computing}, pages 109--119, 2016.

\bibitem{Gupta2010}
Sidhant Gupta, M.S. Reynolds, and S.N. Patel.
\newblock {ElectriSense: single-point sensing using EMI for electrical event
  detection and classification in the home}.
\newblock {\em Proceedings of the 12th ACM international conference on
  Ubiquitous computing}, pages 139--148, 2010.

\bibitem{Campbell:2010:WMS:1864349.1864378}
Tim Campbell, Eric Larson, Gabe Cohn, Jon Froehlich, Ramses Alcaide, and
  Shwetak~N. Patel.
\newblock Wattr: A method for self-powered wireless sensing of water activity
  in the home.
\newblock In {\em Proceedings of the 12th ACM International Conference on
  Ubiquitous Computing}, UbiComp '10, pages 169--172, New York, NY, USA, 2010.
  ACM.

\bibitem{tsai2007usability}
Christopher~C Tsai, Gunny Lee, Fred Raab, Gregory~J Norman, Timothy Sohn,
  William~G Griswold, and Kevin Patrick.
\newblock Usability and feasibility of pmeb: a mobile phone application for
  monitoring real time caloric balance.
\newblock {\em Mobile networks and applications}, 12(2-3):173--184, 2007.

\bibitem{chung2019identifying}
Chia-Fang Chung, Qiaosi Wang, Jessica Schroeder, Allison Cole, Jasmine Zia,
  James Fogarty, and Sean~A Munson.
\newblock Identifying and planning for individualized change: Patient-provider
  collaboration using lightweight food diaries in healthy eating and irritable
  bowel syndrome.
\newblock {\em Proceedings of the ACM on interactive, mobile, wearable and
  ubiquitous technologies}, 3(1):1--27, 2019.

\bibitem{chung2017personal}
Chia-Fang Chung, Elena Agapie, Jessica Schroeder, Sonali Mishra, James Fogarty,
  and Sean~A Munson.
\newblock When personal tracking becomes social: Examining the use of instagram
  for healthy eating.
\newblock In {\em Proceedings of the 2017 CHI Conference on Human Factors in
  Computing Systems}, pages 1674--1687, 2017.

\bibitem{peesapati2010pensieve}
S~Tejaswi Peesapati, Victoria Schwanda, Johnathon Schultz, Matt Lepage, So-yae
  Jeong, and Dan Cosley.
\newblock Pensieve: supporting everyday reminiscence.
\newblock In {\em Proceedings of the SIGCHI Conference on Human Factors in
  Computing Systems}, pages 2027--2036, 2010.

\bibitem{lindqvist2011m}
Janne Lindqvist, Justin Cranshaw, Jason Wiese, Jason Hong, and John Zimmerman.
\newblock I'm the mayor of my house: examining why people use foursquare-a
  social-driven location sharing application.
\newblock In {\em Proceedings of the SIGCHI conference on human factors in
  computing systems}, pages 2409--2418, 2011.

\bibitem{li2010stage}
Ian Li, Anind Dey, and Jodi Forlizzi.
\newblock A stage-based model of personal informatics systems.
\newblock In {\em Proceedings of the SIGCHI conference on human factors in
  computing systems}, pages 557--566. ACM, 2010.

\bibitem{epstein2015lived}
Daniel~A Epstein, An~Ping, James Fogarty, and Sean~A Munson.
\newblock A lived informatics model of personal informatics.
\newblock In {\em Proceedings of the 2015 ACM International Joint Conference on
  Pervasive and Ubiquitous Computing}, pages 731--742. ACM, 2015.

\bibitem{kersten2017personal}
Elisabeth~T Kersten-van Dijk, Joyce~HDM Westerink, Femke Beute, and Wijnand~A
  IJsselsteijn.
\newblock Personal informatics, self-insight, and behavior change: A critical
  review of current literature.
\newblock {\em Human--Computer Interaction}, 32(5-6):268--296, 2017.

\bibitem{schuler1993participatory}
Douglas Schuler and Aki Namioka.
\newblock {\em Participatory design: Principles and practices}.
\newblock CRC Press, 1993.

\bibitem{fischer2016just}
Joel~E Fischer, Andy Crabtree, Tom Rodden, James~A Colley, Enrico Costanza,
  Michael~O Jewell, and Sarvapali~D Ramchurn.
\newblock Just whack it on until it gets hot: Working with iot data in the
  home.
\newblock In {\em Proceedings of the 2016 CHI Conference on Human Factors in
  Computing Systems}, pages 5933--5944. ACM, 2016.

\bibitem{Fischer2017dataWork}
Joel~E. Fischer, Andy Crabtree, James~A. Colley, Tom Rodden, and Enrico
  Costanza.
\newblock {Data Work: How Energy Advisors and Clients Make IoT Data
  Accountable}.
\newblock {\em CSCW 2017}, 26(4-6):597--626.

\bibitem{Browne2011}
Jeffrey Browne, Bongshin Lee, Sheelagh Carpendale, Nathalie Riche, and Timothy
  Sherwood.
\newblock {Data analysis on interactive whiteboards through sketch-based
  interaction}.
\newblock {\em Proceedings of the ACM International Conference on Interactive
  Tabletops and Surfaces, ITS'11}, (December 2014):154--157, 2011.

\bibitem{buxton2007anatomy}
Bill Buxton.
\newblock The anatomy of sketching.
\newblock {\em Sketching User Experiences. Getting the Design Right and the
  Right Design. Morgan Kauffman (2007)}, pages 105--113, 2007.

\bibitem{Lee2013}
Bongshin Lee, Rubaiat~Habib Kazi, and Greg Smith.
\newblock {SketchStory: Telling more engaging stories with data through
  freeform sketching}.
\newblock {\em IEEE Transactions on Visualization and Computer Graphics},
  19(12):2416--2425, 2013.

\bibitem{buxton2010sketching}
Bill Buxton.
\newblock {\em Sketching user experiences: getting the design right and the
  right design}.
\newblock Morgan kaufmann, 2010.

\bibitem{Walny2011}
Jagoda Walny, Sheelagh Carpendale, Nathalie~Henry Riche, Gina Venolia, and
  Philip Fawcett.
\newblock {Visual thinking in action: Visualizations as used on whiteboards}.
\newblock {\em IEEE Transactions on Visualization and Computer Graphics},
  17(12):2508--2517, 2011.

\bibitem{jansen2015opportunities}
Yvonne Jansen, Pierre Dragicevic, Petra Isenberg, Jason Alexander, Abhijit
  Karnik, Johan Kildal, Sriram Subramanian, and Kasper Hornb{\ae}k.
\newblock {Opportunities and Challenges for Data Physicalization}.
\newblock In {\em Proceedings of the 33rd Annual ACM Conference on Human
  Factors in Computing Systems - CHI '15}, pages 3227--3236, New York, New
  York, USA, 2015. ACM Press.

\bibitem{Huron2017}
Samuel Huron, Pauline Gourlet, Uta Hinrichs, Trevor Hogan, and Yvonne Jansen.
\newblock {Let's Get Physical: Promoting Data Physicalization in Workshop
  Formats}.
\newblock {\em Proceedings of the ACM Conference on Designing Interactive
  Systems}, 3(Umr 9217):1409--1422, 2017.

\bibitem{willett2016constructive}
Wesley Willett and Samuel Huron.
\newblock {A Constructive Classroom Exercise for Teaching InfoVis}.
\newblock In {\em {Pedagogy of Data Visualization Workshop at IEEE VIS 2016}},
  Pedagogy of Data Visualization Workshop at IEEE VIS 2016, Baltimore, United
  States, October 2016.

\bibitem{huang2014personal}
Dandan Huang, Melanie Tory, Bon~Adriel Aseniero, Lyn Bartram, Scott Bateman,
  Sheelagh Carpendale, Anthony Tang, and Robert Woodbury.
\newblock Personal visualization and personal visual analytics.
\newblock {\em IEEE Transactions on Visualization and Computer Graphics},
  21(3):420--433, 2014.

\bibitem{choe2017understanding}
Eun~Kyoung Choe, Bongshin Lee, Haining Zhu, Nathalie~Henry Riche, and Dominikus
  Baur.
\newblock Understanding self-reflection: how people reflect on personal data
  through visual data exploration.
\newblock In {\em Proceedings of the 11th EAI International Conference on
  Pervasive Computing Technologies for Healthcare}, pages 173--182. ACM, 2017.

\bibitem{Arias-Hernandez2011}
Richard Arias-Hernandez, Linda~T. Kaastra, Tera~M. Green, and Brian Fisher.
\newblock {Pair Analytics: Capturing Reasoning Processes in Collaborative
  Visual Analytics}.
\newblock In {\em 2011 44th Hawaii International Conference on System
  Sciences}, pages 1--10. IEEE, jan 2011.

\bibitem{trickett2000dipsy}
Susan~B Trickett, Wai-Tat Fu, Chrisitan~D Schunn, and J~Gregory Trafton.
\newblock From dipsy-doodle to streaming motions: Changes in representation in
  the analysis of visual scientific data.
\newblock In {\em Proceedings of the Annual Meeting of the Cognitive Science
  Society}, volume~22, 2000.

\bibitem{dickson2000effects}
Janet Dickson, Jim McLennan, and Mary~M Omodei.
\newblock Effects of concurrent verbalization on a time-critical, dynamic
  decision-making task.
\newblock {\em The Journal of general psychology}, 127(2):217--228, 2000.

\bibitem{wilson1994proper}
Timothy~D Wilson.
\newblock The proper protocol: Validity and completeness of verbal reports.
\newblock {\em Psychological Science}, 5(5):249--252, 1994.

\bibitem{kelley1984iterative}
John~F Kelley.
\newblock An iterative design methodology for user-friendly natural language
  office information applications.
\newblock {\em ACM Transactions on Information Systems (TOIS)}, 2(1):26--41,
  1984.

\bibitem{Kerzner2019}
Ethan Kerzner, Sarah Goodwin, Jason Dykes, Sara Jones, and Miriah Meyer.
\newblock {A Framework for Creative Visualization-Opportunities Workshops}.
\newblock {\em IEEE Transactions on Visualization and Computer Graphics},
  25(1):748--758, jan 2019.

\bibitem{monforte2021pairinterivew}
Javier Monforte and Joan Úbeda Colomer.
\newblock Tinkering with the two-to-one interview: Reflections on the use of
  two interviewers in qualitative constructionist inquiry.
\newblock {\em Methods in Psychology}, 2021.

\bibitem{PRISMS}
{National Institute of Biomedical Imaging Bioengineering}.
\newblock {Pediatric Research Using Integrated Sensor Monitoring Systems},
  2015.

\bibitem{kluyver2016jupyter}
Thomas Kluyver, Benjamin Ragan-Kelley, Fernando P{\'e}rez, Brian~E Granger,
  Matthias Bussonnier, Jonathan Frederic, Kyle Kelley, Jessica~B Hamrick, Jason
  Grout, Sylvain Corlay, et~al.
\newblock Jupyter notebooks-a publishing format for reproducible computational
  workflows.
\newblock In {\em ELPUB}, pages 87--90, 2016.

\bibitem{fonteyn1993description}
Marsha~E Fonteyn, Benjamin Kuipers, and Susan~J Grobe.
\newblock A description of think aloud method and protocol analysis.
\newblock {\em Qualitative health research}, 3(4):430--441, 1993.

\bibitem{lenzner2016cognitive}
Timo Lenzner, Cornelia Neuert, and W~Otto.
\newblock Cognitive pretesting.
\newblock {\em GESIS Survey Guidelines}, page~3, 2016.

\bibitem{presser2004methods}
Stanley Presser, Mick~P Couper, Judith~T Lessler, Elizabeth Martin, Jean
  Martin, Jennifer~M Rothgeb, and Eleanor Singer.
\newblock Methods for testing and evaluating survey questions.
\newblock {\em Public opinion quarterly}, 68(1):109--130, 2004.

\bibitem{kandel2011wrangler}
Sean Kandel, Andreas Paepcke, Joseph Hellerstein, and Jeffrey Heer.
\newblock Wrangler: Interactive visual specification of data transformation
  scripts.
\newblock In {\em Proceedings of the SIGCHI Conference on Human Factors in
  Computing Systems}, pages 3363--3372. ACM, 2011.

\bibitem{rattenbury2017principles}
Tye Rattenbury, Joseph~M Hellerstein, Jeffrey Heer, Sean Kandel, and Connor
  Carreras.
\newblock {\em Principles of data wrangling: Practical techniques for data
  preparation}.
\newblock " O'Reilly Media, Inc.", 2017.

\bibitem{kirsh01work}
David Kirsh.
\newblock The context of work.
\newblock {\em Human–Computer Interaction}, 16(2-4):305--322, 2001.

\bibitem{lee2019correlation}
Sukwon Lee, Bum~Chul Kwon, Jiming Yang, Byung~Cheol Lee, and Sung-Hee Kim.
\newblock The correlation between users’ cognitive characteristics and
  visualization literacy.
\newblock {\em Applied Sciences}, 9(3):488, 2019.

\bibitem{galesic2011graph}
Mirta Galesic and Rocio Garcia-Retamero.
\newblock Graph literacy: a cross-cultural comparison.
\newblock {\em Medical Decision Making}, 31(3):444--457, 2011.

\bibitem{munzner2014visualization}
Tamara Munzner.
\newblock {\em Visualization analysis and design}.
\newblock CRC press, 2014.

\bibitem{iliinsky2011designing}
Noah Iliinsky and Julie Steele.
\newblock {\em Designing data visualizations: Representing informational
  Relationships}.
\newblock " O'Reilly Media, Inc.", 2011.

\bibitem{mccracken1988long}
Grant McCracken.
\newblock {\em The long interview}, volume~13.
\newblock Sage, 1988.

\bibitem{leech2002asking}
Beth~L Leech.
\newblock Asking questions: Techniques for semistructured interviews.
\newblock {\em PS: Political science and politics}, 35(4):665--668, 2002.

\bibitem{smoothedZScore}
Smoothed z score algorithm, stack overflow.
\newblock
  \url{https://stackoverflow.com/questions/22583391/peak-signal-detection-in-realtime-timeseries-data/22640362\#22640362},
  March 25, 2014.
\newblock Accessed: 2017-05-15.

\bibitem{Bentley2013}
Frank Bentley, Konrad Tollmar, Peter Stephenson, Laura Levy, Brian Jones, Scott
  Robertson, Ed~Price, Richard Catrambone, and Jeff Wilson.
\newblock {Health Mashups: Presenting Statistical Patterns between Wellbeing
  Data and Context in Natural Language to Promote Behavior Change}.
\newblock {\em ACM Transactions on Computer-Human Interaction}, 20(5):1--27,
  2013.

\bibitem{jones2015exploring}
Simon~L Jones.
\newblock Exploring correlational information in aggregated quantified self
  data dashboards.
\newblock In {\em Adjunct Proceedings of the 2015 ACM International Joint
  Conference on Pervasive and Ubiquitous Computing and Proceedings of the 2015
  ACM International Symposium on Wearable Computers}, pages 1075--1080. ACM,
  2015.

\bibitem{jia2013asthma}
Chun~E Jia, Hong~Ping Zhang, Yan Lv, Rui Liang, Yun~Qiu Jiang, Heather Powell,
  Juan~Juan Fu, Lei Wang, Peter~Gerard Gibson, and Gang Wang.
\newblock The asthma control test and asthma control questionnaire for
  assessing asthma control: systematic review and meta-analysis.
\newblock {\em Journal of Allergy and Clinical Immunology}, 131(3):695--703,
  2013.

\bibitem{beyer1999contextual}
Hugh Beyer and Karen Holtzblatt.
\newblock Contextual design.
\newblock {\em interactions}, 6(1):32--42, 1999.

\bibitem{card1999readings}
Mackinlay Card.
\newblock {\em Readings in information visualization: using vision to think}.
\newblock Morgan Kaufmann, 1999.

\bibitem{shneiderman2004designing}
Ben Shneiderman.
\newblock Designing for fun: how can we design user interfaces to be more fun?
\newblock {\em interactions}, 11(5):48--50, 2004.

\bibitem{sedlmair2012design}
Michael Sedlmair, Miriah Meyer, and Tamara Munzner.
\newblock Design study methodology: Reflections from the trenches and the
  stacks.
\newblock {\em IEEE transactions on visualization and computer graphics},
  18(12):2431--2440, 2012.

\bibitem{brehmer2013multi}
Matthew Brehmer and Tamara Munzner.
\newblock A multi-level typology of abstract visualization tasks.
\newblock {\em IEEE transactions on visualization and computer graphics},
  19(12):2376--2385, 2013.

\bibitem{slcair-country}
Cimaron Neugebaur.
\newblock Salt lake city has the worse air quality in the nation,
  https://kutv.com/news/local/salt-lake-city-has-the-worst-air-quality-in-the-nation,
  accessed: 2021-10-15.
\newblock Accessed:2021-10-15.

\bibitem{cordeiro2015barriers}
Felicia Cordeiro, Daniel~A Epstein, Edison Thomaz, Elizabeth Bales, Arvind~K
  Jagannathan, Gregory~D Abowd, and James Fogarty.
\newblock Barriers and negative nudges: Exploring challenges in food
  journaling.
\newblock In {\em Proceedings of the 33rd Annual ACM Conference on Human
  Factors in Computing Systems}, pages 1159--1162, 2015.

\bibitem{Li2012}
Ian Li, Anind~K. Dey, and Jodi Forlizzi.
\newblock {Using context to reveal factors that affect physical activity}.
\newblock {\em ACM Transactions on Computer-Human Interaction}, 19(1):1--21,
  2012.

\bibitem{Choe2017semiautomated}
Eun~Kyoung Choe, Saeed Abdullah, Mashfiqui Rabbi, Edison Thomaz, Daniel~A.
  Epstein, Felicia Cordeiro, Matthew Kay, Gregory~D. Abowd, Tanzeem Choudhury,
  James Fogarty, Bongshin Lee, Mark Matthews, and Julie~A. Kientz.
\newblock {Semi-Automated Tracking: A Balanced Approach for Self-Monitoring
  Applications}.
\newblock {\em IEEE Pervasive Computing}, 16(1):74--84, jan 2017.

\bibitem{dasu2003exploratory}
Tamraparni Dasu and Theodore Johnson.
\newblock {\em Exploratory data mining and data cleaning}, volume 479.
\newblock John Wiley \& Sons, 2003.

\bibitem{kennedy2015digital}
Jenny Kennedy, Bjorn Nansen, Michael Arnold, Rowan Wilken, and Martin Gibbs.
\newblock Digital housekeepers and domestic expertise in the networked home.
\newblock {\em Convergence}, 21(4):408--422, 2015.

\bibitem{epstein2020}
Daniel~A. Epstein, Clara Caldeira, Mayara~Costa Figueiredo, Xi~Lu, Lucas~M.
  Silva, Lucretia Williams, Jong~Ho Lee, Qingyang Li, Simran Ahuja, Qiuer Chen,
  Payam Dowlatyari, Craig Hilby, Sazeda Sultana, Elizabeth~V. Eikey, and Yunan
  Chen.
\newblock Mapping and taking stock of the personal informatics literature.
\newblock {\em Proc. ACM Interact. Mob. Wearable Ubiquitous Technol.}, 4(4),
  December 2020.

\bibitem{dahlback1993wizard}
Nils Dahlb{\"a}ck, Arne J{\"o}nsson, and Lars Ahrenberg.
\newblock Wizard of oz studies—why and how.
\newblock {\em Knowledge-based systems}, 6(4):258--266, 1993.

\end{thebibliography}

\end{document}